\documentclass[aps,prb,reprint,superscriptaddress,amsmath,amssymb]{revtex4-2}
\usepackage[dvipdfmx]{graphicx}% Include figure files
\usepackage{dcolumn}% Align table columns on decimal point
\usepackage{bm}% bold math
\usepackage{color}
\begin{document}

%\preprint{}

%\title{Distorted triangular skyrmion lattice in a noncentrosymmetric tetragonal magnet EuNiGe$_3$}
%\title{Helicity reversal in polar tetragonal magnet to form triangular skyrmion lattice}
\title{Single helicity of the triple-$q$ triangular skyrmion lattice state in cubic chiral helimagnet EuPtSi}

\author{Takeshi Matsumura}
\email[]{tmatsu@hiroshima-u.ac.jp}
%\thanks{}
\affiliation{Department of Quantum Matter, ADSE, Hiroshima University, Higashi-Hiroshima 739-8530, Japan}
%\affiliation{Institute for Advanced Materials Research, Hiroshima University, Higashi-Hiroshima 739-8530, Japan}
\author{Chihiro Tabata}
\affiliation{Materials Sciences Research Center, Japan Atomic Energy Agency, Tokai, Ibaraki 319-1195, Japan}
\affiliation{Advanced Science Research Center, Japan Atomic Energy Agency, Tokai, Ibaraki 319-1195, Japan }
\author{Koji Kaneko}
\affiliation{Materials Sciences Research Center, Japan Atomic Energy Agency, Tokai, Ibaraki 319-1195, Japan}
\affiliation{Advanced Science Research Center, Japan Atomic Energy Agency, Tokai, Ibaraki 319-1195, Japan }
\author{Hironori Nakao}
\affiliation{Photon Factory, Institute of Materials Structure Science, High Energy Accelerator Research Organization, Tsukuba, 305-0801, Japan}
\author{Masashi Kakihana}
\affiliation{Faculty of Science, University of the Ryukyus, Nishihara, Okinawa 903-0213, Japan}
\author{Masato Hedo}
\affiliation{Faculty of Science, University of the Ryukyus, Nishihara, Okinawa 903-0213, Japan}
\author{Takao Nakama}
\affiliation{Faculty of Science, University of the Ryukyus, Nishihara, Okinawa 903-0213, Japan}
\author{Yoshichika \={O}nuki}
\affiliation{Faculty of Science, University of the Ryukyus, Nishihara, Okinawa 903-0213, Japan}
\affiliation{RIKEN Center for Emergent Matter Science, Wako, Saitama 351-0198, Japan}


\date{\today}

\begin{abstract}
We investigated the magnetic helicity of the triple-$\bm{q}$ magnetic structure of the triangular skyrmion lattice in the  ``A-phase" of EuPtSi for a magnetic field along the [111] axis by resonant x-ray diffraction using a circularly polarized beam. We show that all three Fourier components of the triple-$\bm{q}$ structure are perpendicular to the respective $\bm{q}$ vectors and have the same helicity. They are connected by the rotation operations about the [111] axis. The helicity is the same as that of the single-$\bm{q}$ helimagnetic phase at low fields, suggesting that the antisymmetric exchange interaction inherent in the chiral structure supports the formation of the triangular skyrmion lattice. We also observe that the helical plane in the helimagnetic phase is tilted to the magnetic field to form a conical structure before the first-order transition to the skyrmion lattice phase.   
\end{abstract}

\maketitle

%\section{Introduction}
\section{Introduction}
A cubic chiral helimagnet EuPtSi, with an ordering temperature of $T_{\text{N}}$=4.0 K, exhibits an emergent magnetic ordered phase in magnetic fields~\cite{Kakihana18,Kakihana19,Takeuchi19,Sakakibara19,Takeuchi20,Sakakibara21,Mishra19}. This ordered phase is called the ``A-phase" since it is reminiscent of the similar phase observed in MnSi just below the ordering temperature and in a finite magnetic field range. 
This interesting magnetic structure was originally clarified in MnSi to exist as a crystallization of spin-swirling particle-like objects composed of three helimagnetic modulation waves, which was named triangular skyrmion lattice (SkL)~\cite{Muhlbauer09,Nagaosa13,Tokura21}. 
In rare-earth EuPtSi, which belongs to the same crystallographic space group $P2_13$ as MnSi, the phase stability is more extended to lower temperatures than that of MnSi.  
This is accompanied by a giant anomalous Hall effect, suggesting an emergent field originating from the formation of a magnetic SkL~\cite{Kakihana18}.

\begin{figure}[t]
\begin{center}
\includegraphics[width=8cm]{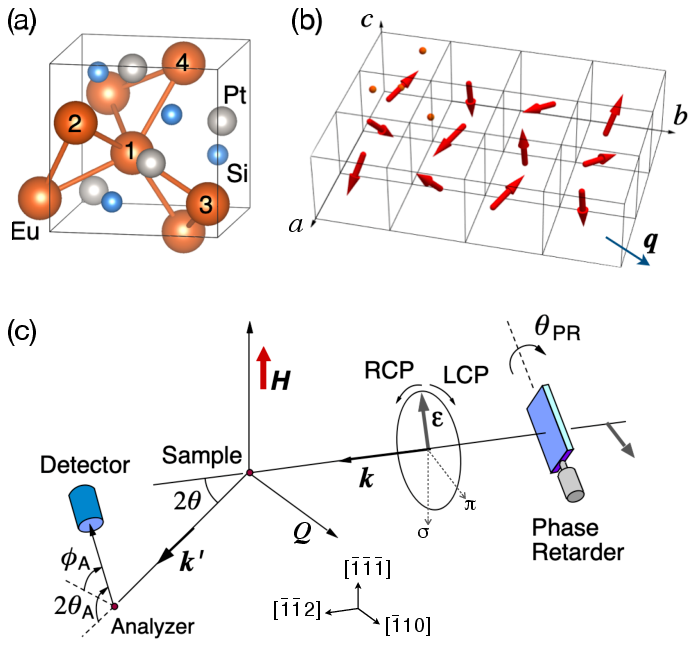} 
\end{center}
\caption{(a) Crystal structure of EuPtSi with four Eu atoms in a unit cell. 
(b) Helical magnetic structure of the zero-field ground state described by $\bm{q}=(0.2, 0.3, 0)$. Only the magnetic moments of Eu-1 are shown. The moments at Eu-2, -3, and -4 are omitted. 
(c) Scattering geometry of the experiment. The phase retarder is removed when linear polarization analysis is performed.  
When $\theta_{\text{PR}}$ scan is performed, the detector is placed directly on the diffracted beam along $\bm{k}^{\prime}$. 
}
\label{fig:fig1}
\end{figure}

The formation of SkL in magnetic fields for $\bm{H} \parallel [111]$ has been demonstrated by the observation of a triple-$\bm{q}$ magnetic order with $\bm{q}_1=(-\delta_3, \delta_1, \delta_2)$,  $\bm{q}_2=(\delta_2, -\delta_3, \delta_1)$, and $\bm{q}_3=(\delta_1, \delta_2, -\delta_3)$ ($\delta_1$=0.09, $\delta_2$=0.20, $\delta_3$=0.29), where $\bm{q}_i \perp \bm{H}$ is realized,  by neutron and resonant x-ray diffraction~\cite{Kaneko19,Tabata19}. 
The single-$\bm{q}$ ordering of the zero-field ground state below $T_{\text{N}}^*$=2.5 K with $\bm{q}=(0.2, 0.3, 0)$ was also established~\cite{Kaneko19,Tabata19}.
The crystal and helimagnetic structures at zero field are shown in Fig. \ref{fig:fig1}(a) and \ref{fig:fig1}(b), respectively.

In chiral magnets without either space inversion or mirror reflection symmetry, the Dzyaloshinskii-Moriya (DM)-type antisymmetric exchange interaction in the form of $\bm{D}_{ij} \cdot (\bm{S}_i \times \bm{S}_j)$ arises, or equivalently, $\bm{D}(\bm{q}) \cdot (\bm{S}_{\bm{q}} \times \bm{S}_{-\bm{q}})$ in the reciprocal space. 
This leads to the selection of a single helicity and lifting of the chiral degeneracy of the helimagnetic spiral. This is actually realized in EuPtSi at zero field and was confirmed by polarized neutron diffraction~\cite{Kaneko19}. 
It should be noted that the short period of the incommensurate spiral is determined by the symmetric exchange interactions of Ruderman-Kittel-Kasuya-Yosida (RKKY) type and the weak antisymmetric exchange interaction lifts the chiral degeneracy.

In the triple-$\bm{q}$ SkL phase of the Bloch type, which is described by a superposition of three helimagnetic waves, the magnetic helicities of the three-component waves must be the same~\cite{Muhlbauer09}. Although the observation of a higher harmonic diffraction peak provides strong evidence for the triple-$\bm{q}$ SkL~\cite{Tabata19}, an experimental observation of the spin-swirling structure in one direction, that is, a direct observation of the single helicity, is necessary to confirm the formation of the SkL.

For this purpose, we employed resonant x-ray diffraction (RXD) with a circularly polarized beam. 
This is a direct observation in the reciprocal space, which is complementary to the real-space observation of spin-swirling structures by Lorentz transmission electron microscopy~\cite{Yu10,Khanh20}.
In this study, we demonstrate that all three Fourier components of the triple-$\bm{q}$ structure in EuPtSi have the same helicity. 
In addition, we show that the helical planes are almost circular and perpendicular to the $\bm{q}$ vector, although they are not the necessary requirements of the symmetry. The single-$\bm{q}$ helimagnetic structure in the low-field phase is also investigated in detail. We show that the helical plane, which is perpendicular to the $\bm{q}$ vector at zero field, is slightly tilted toward the magnetic field direction to form a conical structure to gain the Zeeman energy.

Another interesting aspect of EuPtSi is the geometrical frustration of the $S=7/2$ spins on the three-dimensional network of corner-sharing equilateral triangles, as shown in Fig. \ref{fig:fig1}(a), which is called the trillium lattice~\cite{Hopkinson06,Isakov08,Redpath10}. 
The helimagnetic transition at $T_{\text{N}}$=4.0 K, with $\bm{q}=(0.2, 0.3, \delta)$ ($\delta=0.04$ at $T_{\text{N}}$)~\cite{Kaneko19}, is a first order transition accompanied by a sharp peak in specific heat and a discontinuous magnetization jump, which are superimposed on strong indications of magnetic fluctuations above $T_{\text{N}}$~\cite{Franco17,Sakakibara19,Homma19,Higa21}.
We propose a model helimagnetic structure of the zero-field ground state by considering first and second nearest-neighbor Heisenberg-type exchange interactions.

\section{Experiment}
The RXD experiment was performed at BL-3A of the Photon Factory, KEK, Japan. We used the same EuPtSi single crystal used in Ref. \onlinecite{Tabata19}. The x-ray energy was tuned to 7.614 keV at the resonance of the Eu-$L_2$ absorption edge. 
See Ref. \onlinecite{Tabata19} for the resonant energy spectrum of the magnetic diffraction peak. 
The geometry of the RXD experiment is shown in Fig.~\ref{fig:fig1}(c). 
As in Ref. \onlinecite{Tabata19}, the scattering plane was spanned by the $[\bar{1} 1 0]$ and $[\bar{1} \bar{1} 2]$ axes and the magnetic field was applied along the $[\bar{1} \bar{1} \bar{1}]$ direction.

A circularly polarized beam was obtained using a diamond phase-retarder system. The incident linear polarization is tuned to right-handed circular polarziation (RCP) and left-handed circular polarization (LCP) by manipulating $\Delta \theta_{\text{PR}}=\theta_{\text{PR}} - \theta_{\text{B}}$, where $\theta_{\text{B}}$ is the 111 Bragg angle of the diamond phase-plate. The degrees of circular and linear polarization can be expressed as $P_2 = \sin (\gamma/ \Delta \theta_{\text{PR}})$ ($+1$ for RCP and $-1$ for LCP) and $P_3 = -\cos (\gamma/ \Delta \theta_{\text{PR}})$ ($+1$ for $\sigma$ and $-1$ for $\pi$), respectively, where $\gamma$ is an experimentally determined parameter of the phase plate obtained by analyzing the $\Delta \theta_{\text{PR}}$ dependence of the intensity of the $(\bar{2}, 2, 0)$ fundamental reflection, as explained in the Appendix.
The $\Delta \theta_{\text{PR}}$-scan of the magnetic Bragg-peak intensity is sensitive for determining the helicity of the Fourier component. 
A linear polarization analysis of the diffracted x ray for the $\pi$-polarized incident beam was also performed to determine the Fourier component more precisely. We used the 006 Bragg reflection of a pyrolytic graphite (PG) analyzer crystal. The intensity variation was measured as a function of the detector angle $(\phi_{\text{A}})$ measured from the horizontal scattering plane. 
This analysis is more suitable for estimating the ratio between the horizontal and vertical components, from which we can estimate the ellipticity of the helical plane.

\section{Results and Analysis}
\subsection{helical magnetic structure at zero field}
In the helical magnetic phase at zero field, by combining the $\Delta \theta_{\text{PR}}$-scans and linear polarization analysis ($\phi_{\text{A}}$ scans), we confirmed that the helical plane is almost perpendicular to the $\bm{q}$ vector and is almost circular. 
In the single-$\bm{q}$ structure at zero field, the magnetic moment of Eu-$\alpha$ ($\alpha=1\sim 4$) in the $l$-th unit cell at $\bm{r}_l$ is generally expressed as 
\begin{equation}
\bm{\mu}_{\alpha,l} = \bm{m}_{\bm{q},\alpha} e^{i \bm{q}\cdot\bm{r}_l } + \bm{m}_{\bm{q},\alpha}^* e^{-i \bm{q}\cdot\bm{r}_l } 
\label{eq:1}
\end{equation}
using the Fourier component $\bm{m}_{\bm{q},\alpha}$ consisting of real and imaginary parts to express the spiral structure.  
The $E1$ resonant scattering amplitude for the magnetic dipole order is proportional to 
$(\bm{\varepsilon}^{\prime} \times \bm{\varepsilon} ) \cdot \bm{F}_{\text{M}}$, where 
\begin{equation}
\bm{F}_{\text{M}} = \sum_{l,\alpha} \bm{\mu}_{\alpha,l} e^{-i\bm{Q}\cdot (\bm{r}_l + \bm{d}_{\alpha} )}
\label{eq:2}
\end{equation}
is the magnetic structure factor at the scattering vector $\bm{Q}=\bm{k}'-\bm{k}$. 
$\bm{d}_{\alpha}$ represents the atomic position of Eu-$\alpha$ in the unit cell.

\begin{figure}[t]
\begin{center}
\includegraphics[width=8.5cm]{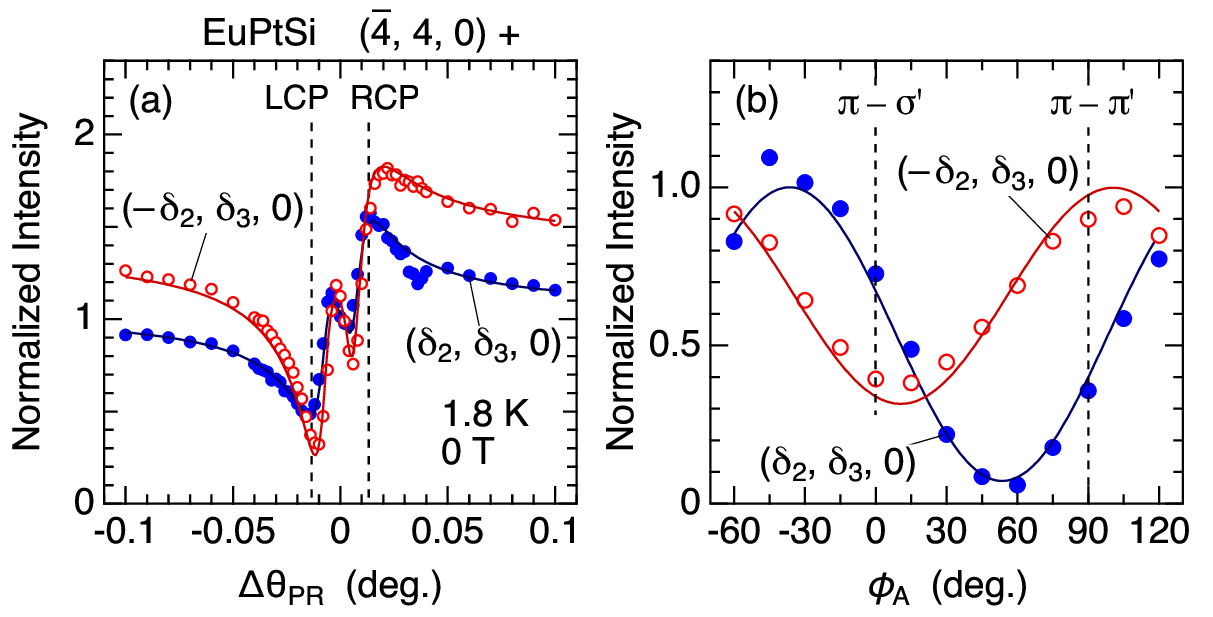} 
\end{center}
\caption{(a) $\Delta \theta_{\text{PR}}$ dependences of the peak intensity at $(\bar{4}, 4, 0)$ + $(\delta_2, \delta_3, 0)$ and $(-\delta_2, \delta_3, 0)$ in the helical phase at 0 T and 1.8 K, where $(\delta_2, \delta_3)=(0.2, 0.3)$. 
The x-ray energy is 7.614 keV at resonance. 
The intensities are normalized at $\Delta \theta_{\text{PR}}=0$, where the raw intensity is 720 cps for $(\delta_2, \delta_3, 0)$ and 980 cps for $(-\delta_2, \delta_3, 0)$. 
The solid lines are the intensities calculated by assuming that the $\bm{m}_{\bm{q}}$ vector is circular, counterclockwise, and perpendicular to the $\bm{q}$ vector. 
(b) Linear polarization analysis of the helimagnetic Bragg peak. 
The solid lines represent calculations assuming the same helimagnetic structure. }
\label{fig:FigPRPOL0T}
\end{figure}

Examples of data analyses and comparisons with the calculated intensity curves are shown in Fig.~\ref{fig:FigPRPOL0T}(a) and \ref{fig:FigPRPOL0T}(b) for $\bm{q}=(-0.2, 0.3, 0)$ and $\bm{q}=(0.2, 0.3, 0)$. 
By combining the $\Delta \theta_{\text{PR}}$-scans and $\phi_{\text{A}}$-scans, we obtain the Fourier components; 
$\bm{m}_{\bm{q},\alpha} = (-3, \pm 2, 3.6 i) e^{i\phi_{\alpha}} $ for $\bm{q}=(\pm 0.2, 0.3, 0)$, 
$\bm{m}_{\bm{q},\alpha} = (3.6 i, -3, \pm 2)  e^{i\phi_{\alpha}} $ for $\bm{q}=(0, \pm 0.2, 0.3)$, and 
$\bm{m}_{\bm{q},\alpha} = (\pm 2, 3.6 i, -3)  e^{i\phi_{\alpha}} $ for $\bm{q}=(0.3, 0, \pm 0.2)$. 
Note that the phases of the Fourier components cannot be obtained from the present experimental data. 
Although this does not affect the analysis, the relative angles between neighboring Eu moments in the real space remain unknown. 
For this reason, we omitted the magnetic moments of Eu-2, 3, 4 in Fig.~\ref{fig:fig1}(b).

In the real space, the magnetic moments rotate counterclockwise when propagating along the $\bm{q}$ vector. 
This result is the same as that obtained by polarized neutron diffraction~\cite{Kaneko19}. 
We can also conclude that the magnetic moments rotate along a circular trajectory perpendicular to $\bm{q}$. 
It should be noted that this is not a symmetry requirement because the direction of the $\bm{m}_{\bm{q}}$ vector has no symmetry restriction according to the irreducible representation for this low-symmetric $\bm{q}$ vector. 
It is also not required that the helical plane to be perpendicular to the $\bm{q}$ vector and circular; i.e., 
the $\bm{D}(\bm{q})$ vector in the reciprocal space does not need to be parallel to $\bm{q}$~\cite{Yambe22}. 
However, the resultant structure suggests that $\bm{D}(\bm{q})$ is parallel to $\bm{q}$ and the energy gain by the exchange interaction is maximized by taking this helical structure. 
Because the crystal field anisotropy for the $S=7/2$ ($L=0$) state of Eu$^{2+}$ is negligible, as inferred from the isotropic magnetic susceptibility above $T_{\text{N}}$~\cite{Sakakibara19,Sakakibara21}, the helical plane is determined to be perpendicular to $\bm{q}$ presumably by the DM-type antisymmetric exchange term, or by the anisotropic exchange term, both originating from antisymmetric spin-orbit interaction inherent in a noncentrosymmetric metallic system~\cite{Hayami18,Hayami21b}.

\subsection{helical magnetic structure at 0.8 T}
\begin{figure}[t]
\begin{center}
\includegraphics[width=8.5cm]{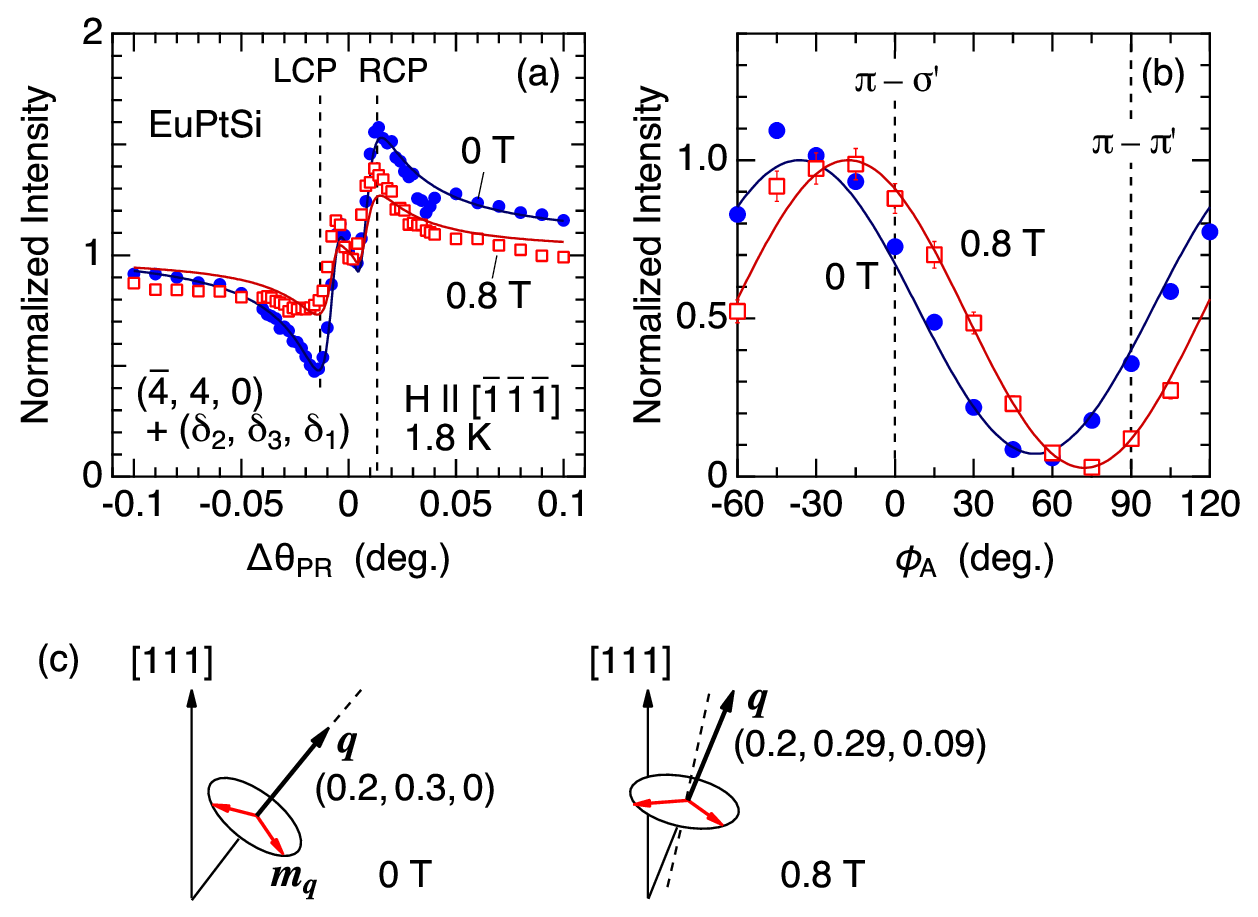} 
\end{center}
\caption{(a) $\Delta \theta_{\text{PR}}$ dependences of the peak intensity at $(\bar{4}, 4, 0)$ + $(\delta_2, \delta_3, \delta_1)$ in the helical phase at 0 T and 0.8 T, where $(\delta_2, \delta_3, \delta_1)=(0.2, 0.29, 0.09)$ at 0.8 T. 
The intensities are normalized at $\Delta \theta_{\text{PR}}=0$, where the raw intensity is 720 cps at 0 T and 840 cps at 0.8 T.  
The solid lines represent the calculated intensities as described in the text. 
(b) Linear polarization analysis of the Bragg peak at 0 T and 0.8 T. The data at 0 T are the same as those in Fig.~\ref{fig:FigPRPOL0T}. 
(c) Relation between the $\bm{q}$ vector, helical plane spanned by the real and imaginary parts of $\bm{m}_{\bm{q}}$, and the [111] axis. The dashed line represents the normal to the helical plane. }
\label{fig:FigPRPOL08T}
\end{figure}

In magnetic fields along $[\bar{1} \bar{1} \bar{1}]$, the helical magnetic domain with the propagation vector $\bm{q}=(\delta_2, \delta_3, \delta_1)$ is selected, where $\delta_2=0.2$ and $\delta_3=0.3$ are nearly constant and $\delta_1$ increases from zero to $\sim 0.09$ at 0.8 T just before the transition to the SkL phase~\cite{Tabata19}. 
The results of the $\Delta \theta_{\text{PR}}$-scan and the $\phi_{\text{A}}$-scan for the helical Bragg peak at $\bm{q}=(0.2, 0.29, 0.09)$ at 0.8 T are shown in Fig.~\ref{fig:FigPRPOL08T}(a) and \ref{fig:FigPRPOL08T}(b), respectively, which are compared with the data at 0 T. 
In the $\Delta \theta_{\text{PR}}$-scan, the decrease and increase in intensity at the LCP and RCP positions, respectively, are less significant than those at 0 T. 
This indicates that the helical plane is tilted to the vertical field direction; in other words, the spiral component parallel to the horizontal scattering plane (conical component) increases. In a perfectly conical structure in which only the horizontal spiral component and uniform magnetization along the vertical field direction exist, the $\Delta \theta_{\text{PR}}$ dependence is expected to be flat.

In the $\phi_{\text{A}}$-scan shown in Fig.~\ref{fig:FigPRPOL08T}(b), the maximum and minimum positions are shifted by $\sim 25^{\circ}$. These two datasets of $\Delta \theta_{\text{PR}}$ and $\phi_{\text{A}}$ scans at 0.8 T can be explained by assuming that the helical plane is more tilted to the vertical field direction by $7\pm 1^{\circ}$ from the position perpendicular to $\bm{q}=(0.2, 0.29, 0.09)$. This process is illustrated in Fig.~\ref{fig:FigPRPOL08T}(c) and the calculated intensities are shown by the solid lines in Fig.~\ref{fig:FigPRPOL08T}(a) and \ref{fig:FigPRPOL08T}(b), which explain the data well. 
These results show that the increase of $\delta_1$ by applying a magnetic field is to gain the Zeeman energy by increasing the horizontal helical component, which is perpendicular to the vertical magnetic field. 
Therefore, the increase in $\delta_1$ indicates that the helical plane and propagation vector $\bm{q}$ are energetically coupled so that they prefer to be perpendicular to each other. At the same time, because there is no such symmetry restriction, they can be decoupled, which leads to the slight tilt of the helical plane by $\sim 7^{\circ}$ that is more perpendicular to the magnetic field, as indicated by the dashed line in Fig.~\ref{fig:FigPRPOL08T}(c).

\subsection{SkL phase}
\begin{figure*}[t]
\begin{center}
\includegraphics[width=15cm]{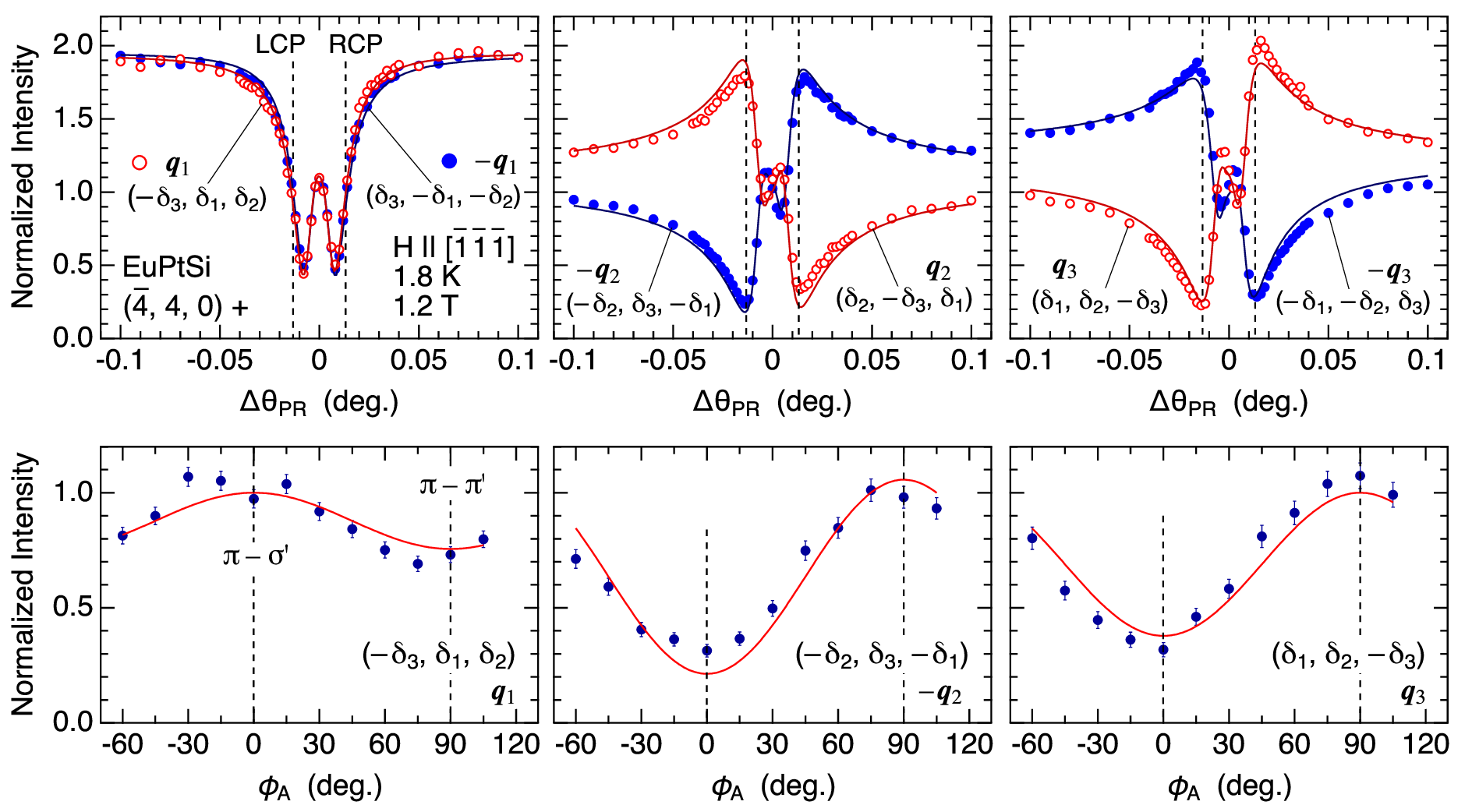} 
\end{center}
\caption{(top panels) $\Delta \theta_{\text{PR}}$ dependences of the three constituent $\bm{q}$ vectors in the SkL phase of EuPtSi for $H \parallel [\bar{1} \bar{1} \bar{1}]$. 
The intensities are normalized at $\Delta \theta_{\text{PR}}=0$, where the raw intensity is 1800 ($\bm{q}_1$), 1010 ($-\bm{q}_1$), 1030 ($\bm{q}_2$), 1850 ($-\bm{q}_2$), 1410 ($\bm{q}_3$), and 1120 ($-\bm{q}_3$) cps. 
The solid lines are the intensities calculated by assuming that the $\bm{m}_{\bm{q}}$ vectors are almost circular ($\sim 12$ \% compressed along the [111] direction), counterclockwise, and perpendicular to the $\bm{q}$ vector. 
$\delta_1$=0.09, $\delta_2$=0.2, and $\delta_3$=0.29. 
(bottom panels) Linear polarization analysis of the corresponding Bragg peaks in the top panels. 
The solid lines are the calculations assuming the same helimagnetic structure. 
}
\label{fig:FigPRPOL12T}
\end{figure*}

In the ``A-phase''  between 0.9 T and 1.3 T, a triple-$\bm{q}$ magnetic structure is realized. The first harmonic structure is expressed as the sum of three $\bm{q}$-components, which is written in the form of Eq.~(\ref{eq:1}),  
\begin{equation}
\bm{\mu}_{\alpha,l} = \sum_{n=1}^3 \{\bm{m}_{\bm{q}_n,\alpha} e^{i (\bm{q}_n\cdot\bm{r}_l + \varphi_n) } + \bm{m}_{\bm{q}_n,\alpha}^* e^{-i (\bm{q}_n \cdot\bm{r}_l  + \varphi_n) } \}\;,
\label{eq:3}
\end{equation}
where $\varphi_n$ represents the phase of each component. 

Top panels of Fig.~\ref{fig:FigPRPOL12T} shows the $\Delta \theta_{\text{PR}}$ dependences for the three constituent $\bm{q}$ vectors in the triple-$\bm{q}$ SkL phase in a magnetic field of 1.2 T $\parallel [\bar{1}\bar{1}\bar{1}]$. The three $\bm{q}$-vectors satisfy the relationship $\bm{q}_1 + \bm{q}_2 + \bm{q}_3=0$. 
As shown by the solid lines, all the data can be explained by assuming the helimagnetic Fourier components of $\bm{m}_{\bm{q}1}$, $\bm{m}_{\bm{q}2}$, and $\bm{m}_{\bm{q}3}$, being almost circular, counter clockwise, and perpendicular to $\bm{q}_1$,  $\bm{q}_2$, and  $\bm{q}_3$, respectively, in the same manner as for the helical structure at zero field. The linear polarization analysis shows that the helical plane is slightly ($\sim 12$ \%) compressed along the [111] field direction. 
To be specific, $\bm{m}_{\bm{q}}$ used in the calculation is expressed as $\bm{v}_{\bm{q}}+i\bm{h}_{\bm{q}}$, where $\bm{v}_{\bm{q}}$ is the vertical component parallel to [111] and $\bm{h}_{\bm{q}}=(\bm{q}/q) \times 1.16\bm{v}_{\bm{q}}$ is the horizontal component. 
%Note that the relative phases between $\bm{m}_{\bm{q}1}$, $\bm{m}_{\bm{q}2}$, and $\bm{m}_{\bm{q}3}$ in Eq.~(\ref{eq:3}) cannot be deduced from the analysis of the present experiment. 

\begin{figure}[t]
\begin{center}
\includegraphics[width=8.5cm]{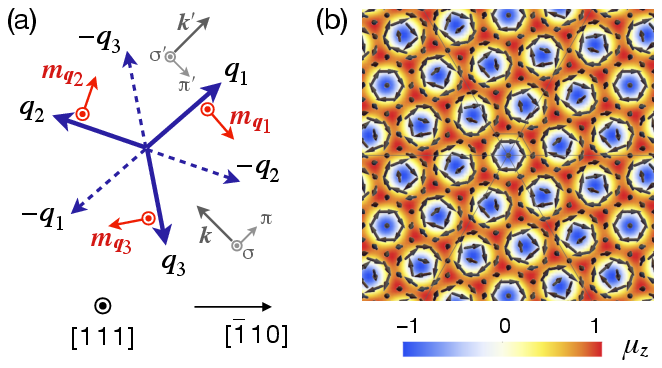} 
\end{center}
\caption{
(a) Three $\bm{q}$ vectors and the corresponding magnetic Fourier components $\bm{m}_{\bm{q}}$, consisting of vertical ($\bm{v}_{\bm{q}} \parallel [111]$) and horizontal ($\bm{h}_{\bm{q}} \perp [111]$) components, in the SkL phase of EuPtSi. The x-ray wave vectors and the polarization vectors represent the geometry for the $(\bar{4}, 4, 0)+\bm{q}$ reflections. 
(b) Top view of the schematic magnetic structure of the Eu-1 atoms in the A-phase. Three layers of Eu-1 atoms are superimposed on each other. 
The phases satisfy $\varphi_1 + \varphi_2 + \varphi_3 = \pi$. The uniform magnetization along the $z$-axis is set to 0.25. The mean magnitude of the magnetic moment $\langle \mu \rangle$ is 0.63, and the standard deviation $\sigma_{\mu}$ is 0.053. 
}
\label{fig:Fig_SkL}
\end{figure}

The three $\Delta \theta_{\text{PR}}$ dependences exhibit different behaviors, as shown in Fig.~\ref{fig:FigPRPOL12T}. 
This is simply due to the geometrical factor of the $E1$ resonant scattering from magnetic dipole moments, which is expressed as $(\bm{\varepsilon}^{\prime} \times \bm{\varepsilon})\cdot \bm{m}_{\bm{q}}$.  
The geometry of the three $\bm{q}$-vectors in the SkL phase for $H \parallel [111]$ (and for  $H \parallel [\bar{1}\bar{1}\bar{1}]$) is shown in Fig.~\ref{fig:Fig_SkL}(a). 
For the $(\bar{4}, 4, 0)+\bm{q}$ reflections here, the $2\theta$ angles are close to $90^{\circ}$ ($97.7^{\circ}$ for $\bm{q}_1$, $84.5^{\circ}$ for $\bm{q}_2$ and $93.3^{\circ}$ for $\bm{q}_3$). 
From Eq.~(\ref{eq:CrossSec1}), the circular polarization ($P_2$) dependent term arises from $F_{\pi\pi'}^*F_{\sigma\pi'}$, because $F_{\sigma\sigma'}=0$ for  magnetic scattering. 
$F_{\pi\pi'}$ arises from $(\bm{\varepsilon}^{\prime}_{\pi} \times \bm{\varepsilon}_{\pi})\cdot \bm{v}_{\bm{q}}$, which is common to all $\bm{q}$ vectors. 
However, $F_{\sigma\pi'}$ arises from $(\bm{\varepsilon}^{\prime}_{\pi} \times \bm{\varepsilon}_{\sigma})\cdot \bm{h}_{\bm{q}}$, where $(\bm{\varepsilon}^{\prime}_{\pi} \times \bm{\varepsilon}_{\sigma})$ is parallel to $\bm{k}^{\prime}$ and is almost perpendicular to $\bm{h}_{\bm{q}1}$. 
This is the reason for the weak $P_2$ dependence in the $(\bar{4}, 4, 0)\pm\bm{q}_1$ reflections and the clear $P_2$ dependences in the $(\bar{4}, 4, 0)\pm\bm{q}_{2,3}$ reflections. 
In contrast, $(\bm{\varepsilon}^{\prime}_{\sigma} \times \bm{\varepsilon}_{\pi})$ is parallel to $\bm{k}$ and is almost parallel to $\bm{h}_{\bm{q}1}$. 
This is the reason for the strong (weak) $\pi$-$\sigma'$ intensity in the linear polarization analysis for $(\bar{4}, 4, 0)+\bm{q}_1$ ($-\bm{q}_2, +\bm{q}_3$). 
This also provides a reason for the longer horizontal component compared with the vertical component.

Figure \ref{fig:Fig_SkL}(b) shows a real-space image of the magnetic structure of Eu-1 atoms viewed from the [111] axis. Three successive (111) planes of Eu-1 atoms are superimposed (1st, 3rd, and 5th layers, as shown in Fig. 12 of Ref.~\onlinecite{Kakihana19}). 
The distance between the skyrmion cores is $\sim 19.9$ \AA. 
Because the phase relations among the three Fourier components are unknown in our diffraction experiment, it is necessary to assume the phases to draw this real-space image. We then set the phases such that the magnetic moment at the origin in Fig. \ref{fig:Fig_SkL}(b) points opposite to the applied field, i.e., $\varphi_1 + \varphi_2 + \varphi_3 = \pi$. A uniform magnetization of 0.25 along the $z$-axis is added to the modulation of Eq.~(\ref{eq:3}) with a maximum amplitude of 1. 
The image thus obtained is consistent with the SkL structure obtained theoretically by numerical simulation~\cite{Hayami21}.
The theory considers the RKKY-type symmetric exchange interaction up to higher-order terms, which is considered to be the origin of the stabilization of the triple-$\bm{q}$ structure~\cite{Hayami17}.

The results of the helicity measurement in Fig.~\ref{fig:FigPRPOL12T} clearly show that the three constituent helimagnetic waves of the triangular SkL in EuPtSi have the same magnetic helicity. They are counterclockwise when propagating along the $\bm{q}$ vector, which is the same as the helicity at zero field. This provides direct evidence in the reciprocal space for the formation of Bloch-type SkL in EuPtSi. 
The single helicity of the triple-$\bm{q}$ structure was also reproduced theoretically by considering the DM-type exchange interaction~\cite{Hayami21}. 
This means that the DM-type exchange interaction inherent in the chiral structure supports the formation of the triangular SkL. 
Simultaneously, we note that the driving force for the formation of the triple-$\bm{q}$ structure is probably the higher-order term in the RKKY interaction and not the DM-type exchange interaction~\cite{Hayami21,Hayami17}. There is even a case where the original helicity at zero field is reversed when a SkL is formed in a magnetic field~\cite{Matsumura23}.

It should be noted that the other magnetic moments of Eu-2, -3, and -4 are omitted in Fig.~\ref{fig:Fig_SkL}(b). 
As in the case of the helical phase at $H=0$, this is because we have no experimental information on the phase relationship or the relative angles among the four Eu moments in a unit cell. 
Because the relative angles should be associated with the geometrical frustration inherent in the trillium lattice structure of Eu, this is another important knowledge to be clarified in the future by more detailed structural analysis. 
However, the triple-$\bm{q}$ SkL in Fig.~\ref{fig:Fig_SkL}(b) should basically be a two-dimensional triangular lattice of skyrmion tubes extending along the [111] axis. Therefore, it is reasonable to consider that the magnetic moments of Eu-2, -3, and -4 are incorporated in the same manner in the SkL structure shown in Fig.~\ref{fig:Fig_SkL}(b). 
Antiferromagnetic coupling among the four Eu moments in the SkL, which would reduce the topological Hall effect, is unlikely.

In a Bloch-type triangular SkL, the magnetic helicities of the three constituent helimagnetic waves must be the same to produce a spin-swirling structure rotating in a specified direction. To experimentally prove this, real-space observation by Lorentz transmission microscopy is a straightforward method. However, it is often difficult to observe short-period SkLs in rare-earth systems, such as in the present case of EuPtSi. 
The present method of RXD utilizing a circularly polarized beam, combined with a linear polarization analysis, provides a direct observation in the reciprocal space, which can be a complementary method. A high spatial resolution and the ability to determine the Fourier components ($\bm{m}_{\bm{q}}$) are significant advantages. However, with respect to the relative phase relation of $\bm{m}_{\bm{q}}$ between different $\bm{q}$ components and different Eu atoms, it is difficult to determine from RXD analysis alone.

\section{Discussion}
\subsection{Other possibilities of the SkL structure}
\begin{figure}[t]
\begin{center}
\includegraphics[width=8.5cm]{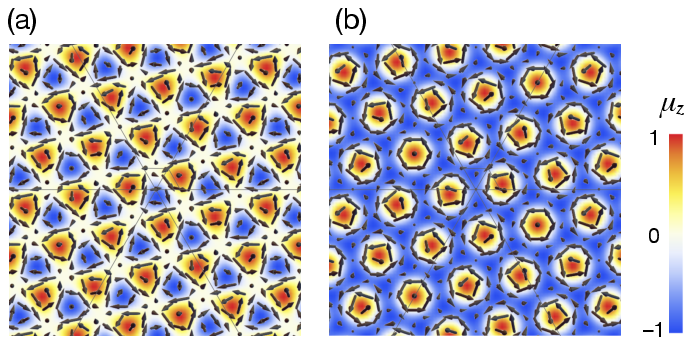} 
\end{center}
\caption{Other possibilities of the SkL structure in the A-phase with different phases characterized by $\tilde{\varphi}=\varphi_1 +  \varphi_2 + \varphi_3$. The uniform magnetic moment along the $z$-axis is set to 0.25. 
(a) $\tilde{\varphi} = \pi/2$. $\langle \mu \rangle = 0.59$, $\sigma_{\mu} = 0.23$. (b) $\tilde{\varphi} = 0$. $\langle \mu \rangle = 0.54$, $\sigma_{\mu} = 0.32$. 
}
\label{fig:SkL2}
\end{figure}

In Fig.~\ref{fig:Fig_SkL}(b), the phases of the three Fourier components are selected so that the center of the skyrmion points opposite to the external magnetic field. If we change the phase relations, different structures are obtained, which are characterized by $\tilde{\varphi}=\varphi_1 +  \varphi_2 + \varphi_3$~\cite{Shimizu22}. Fig.~\ref{fig:Fig_SkL}(b) corresponds to $\tilde{\varphi}=\pi$. 
Although it is difficult to determine this phase relation in this experiment on limited number of reflections, let us discuss the possibilities for alternative structures. 
For instance, by setting $\tilde{\varphi}=\pi/2$ we obtain a structure in which half skyrmions with opposite signs are ordered alternately, as shown in Fig.~\ref{fig:SkL2}(a). Setting $\tilde{\varphi}=0$ results in a SkL where the skyrmion center points towards the magnetic field, while the periphery points opposite to the field, as shown in Fig.~\ref{fig:SkL2}(b). Distinguishing these differences is difficult in diffraction experiments. 
However, it can be concluded from the following discussion that the structure shown in Fig.~\ref{fig:Fig_SkL}(b) is the most plausible. 

When the modulated magnetic structure in Eq.~(\ref{eq:3}) is superimposed onto a uniform magnetization, the calculated magnetic moments at each Eu site generally have unequal magnitudes. For example, in Fig.~\ref{fig:Fig_SkL}(b), where a modulation with a maximum amplitude of 1 is superimposed onto a uniform magnetization of 0.25, the mean magnitude $\langle \mu \rangle$ is 0.63 and the standard deviation $\sigma_{\mu}$ is 0.053, only $\sim 8$ \% of the mean value. 
Notably, this structure yields a minimum $\sigma_{\mu}$. 
In contrast, $\sigma_{\mu}$ for the $\tilde{\varphi} = \pi/2$ case in Fig.~\ref{fig:SkL2}(a) is as high as $\sim 40$ \% of the mean value. It increases to $\sim 59$ \% for the $\tilde{\varphi} = 0$ case in Fig.~\ref{fig:SkL2}(b). 
Considering that such a significant dispersion is unlikely to arise in an ordered state of $S=7/2$ spins of Eu, it seems reasonable to conclude that the SkL structure shown in Fig.~\ref{fig:Fig_SkL}(b) is realized.

Furthermore, the validity of Fig.~\ref{fig:Fig_SkL}(b) can be inferred from the intensity of higher-order reflections such as $\bm{q}_2 - \bm{q}_3$. 
The higher-order reflections do not occur from Eq.~(\ref{eq:3}) only. It is necessary to add some modifications. 
One method is to fix the spin orientations obtained from Eq.~(\ref{eq:3}), and equalize the magnitudes of the moments. 
The difference then gives rise to higher-order terms. 
If we calculate the magnetic structure factor $F_{\text{M}}(\bm{q}_2 - \bm{q}_3)$ for the higher-order reflection at $\bm{q}_2 - \bm{q}_3$ and compare it with the structure factor $F_{\text{M}}(\bm{q}_2)$ of the primary reflection at $\bm{q}_2$, the scattering intensity ratio is estimated to be $|F_{\text{M}}(\bm{q}_2 - \bm{q}_3)/F_{\text{M}}(\bm{q}_2)|^2 \sim 0.0012$. This is comparable to the experimental value of 0.002~\cite{Tabata19}. 
By adjusting $\tilde{\varphi}$ to $0.9\pi$, the calculated intensity ratio agrees with the experimental value. With such a small adjustment, however, the visual structure hardly differs from that shown in Fig.~\ref{fig:Fig_SkL}(b). 
In contrast, applying similar calculations to the cases of Fig.~\ref{fig:SkL2}(a) and (b), the calculated intensity ratios are 0.02 and 0.14, respectively. 
This implies that higher-order reflections should appear at a much larger intensity than that observed. Hence, Fig.~\ref{fig:Fig_SkL}(b) can be considered the most plausible structure. 

\subsection{Helimagnetic structure at zero field }
We omitted the magnetic moments of Eu-2, 3, 4 in Fig.~\ref{fig:fig1}(b) for the single-$\bm{q}$ helical order at zero field, because the relative phases of $\bm{m}_{\bm{q},\alpha}$ for different Eu atoms ($\alpha=1 \sim 4$) have not been determined. 
To discuss the possible helimagnetic structure, let us calculate the classical Heisenberg-type exchange energy, which is expressed by 
\begin{eqnarray}
\mathcal{H} &=& - \sum_{\langle i,j \rangle} J_{ij} \bm{S}_i \cdot \bm{S}_j  \label{eq:4} \\
&=& \frac{2}{N} \sum_{\bm{q}} |\bm{S}_{\bm{q}}^{(\alpha)}\cdot\bm{S}_{-\bm{q}}^{(\beta)}| \hat{s}_{\bm{q}} J_{\bm{q}}^{\alpha\beta} \hat{s}_{-\bm{q}}\;.
\label{eq:5}
\end{eqnarray}
$\hat{s}_{\bm{q}}=(s_{\bm{q}}^{(1)}, s_{\bm{q}}^{(2)}, s_{\bm{q}}^{(3)}, s_{\bm{q}}^{(4)})$ 
represents the phase factors of the Fourier transform $\bm{S}_{\bm{q}}^{(\alpha)} = \sum_j \bm{S}_j  \exp (-i \bm{q}\cdot\bm{r}_j^{(\alpha)})$.  
$J_{\bm{q}}^{\alpha\beta}$ is a $4 \times 4$ matrix ($\alpha,\beta=1\sim 4$)~\cite{Hopkinson06}. 
$J_{ij}>0$ corresponds to ferromagnetic interaction. 

\begin{figure}[t]
\begin{center}
\includegraphics[width=8.5cm]{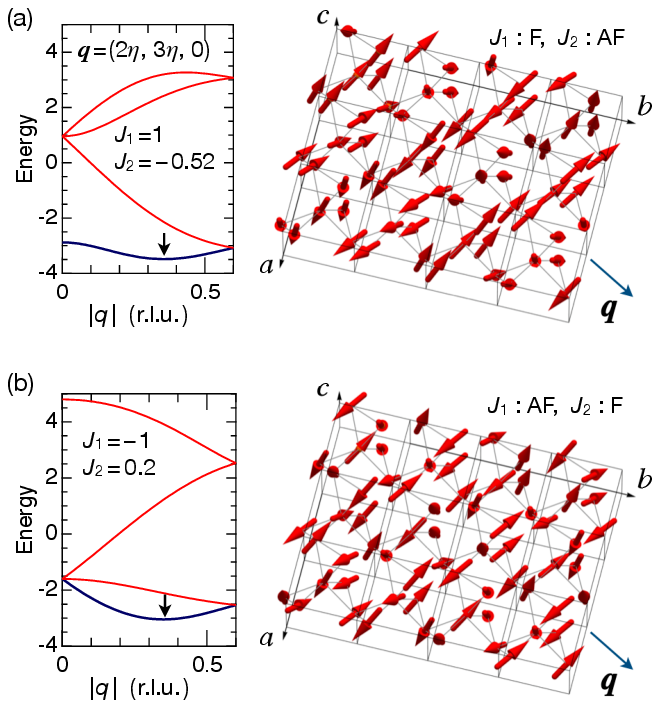} 
\end{center}
\caption{(a) Left: Energy eigenvalues of $J_{\bm{q}}$ for $(J_1, J_2)=(1, -0.52)$ along $\bm{q}=(2\eta, 3\eta, 0)$ as a function of $|q|$ ($\eta=0 \sim 1/6$). 
Right: A Model of helimagnetic structure for $\bm{q}=(0.2, 0.3, 0)$ with ferromagnetic $J_1$ and antiferromagnetic $J_2$.  
(b) Left: Energy eigenvalues of $J_{\bm{q}}$ for $(J_1, J_2)=(-1, 0.2)$. 
Right: A model of helimagnetic structure for $\bm{q}=(0.2, 0.3, 0)$ with antiferromagnetic $J_1$ and ferromagnetic $J_2$.  
}
\label{fig:Helical}
\end{figure}

Let us consider two cases. One is the ferromagnetic (F) nearest-neighbor interaction ($J_1 > 0$) and the other is the antiferromagnetic (AF) nearest-neighbor interaction ($J_1 < 0$). By introducing second nearest neighbor interaction with the opposite sign, $J_2 < 0$ (AF) for the former and $J_2 > 0$ (F) for the latter case, respectively, we can adjust the minimum energy to come midway in the Brillouin zone. Examples of eigenvalues of the $J_{\bm{q}}$ matrix are shown in Fig.~\ref{fig:Helical}(a) for $(J_1, J_2)=(1, -0.52)$ and in Fig.~\ref{fig:Helical}(b) for $(J_1, J_2)=(-1, 0.2)$. The values of $J_2$ are chosen such that the minimum comes at $|q|\sim 0.36$ as indicated by the arrow, which corresponds to $\bm{q}=(0.2, 0.3, 0)$. 

The four eigenstates at each $\bm{q}$ position represent the relative phase (or angle) relationships among the four Eu spins in the unit cell. A simple case is for $\bm{q}=(0,0,0)$. The singlet corresponds to the ferromagnetic arrangement, in which the four spins have the same phase, that is, they are oriented in the same direction. Therefore, the energy of the singlet is lowest (highest) when $J_1 > 0$ ($J_1 < 0$). 
The triplet corresponds to antiferromagnetic arrangements, in which the sum of the four spins is zero. 
Although the precise phase relation becomes more complex at finite $\bm{q}$ positions, the basic phase relation described above does not change. The inclusion of additional exchange parameters also does not seriously affect this qualitative discussion. 

Another important aspect is that the energy minimum and its $|q|$ value of $\sim 0.36$ are almost isotropic, as has generally been studied theoretically~\cite{Hopkinson06,Isakov08}. 
The minimum in the Heisenberg-type exchange energy is realized by $|q| \sim 0.36$ and hardly depends on the direction of $\bm{q}$.
The true minimum of the exchange energy of the RKKY interaction in EuPtSi is at $(\pm 0.2, \pm 0.3, \pm 0.04)$ and its cyclic permutations, which correspond to the peaks in $\chi(\bm{q})$. The 12 helimagnetic structures are degenerate, and other close correlations with $|q| \sim 0.36$ are expected to give rise to a large fluctuation near the ordering temperature. 
Diffuse scattering observed in MnSi provides a good reference~\cite{Janoschek13,Pappas17}. 
When the $\bm{q}$ vector jumps at 0.9 T from $(\delta_2, \delta_3, \delta_1)=(0.2, 0.29, 0.09)$ in the helical phase to $(\delta_2, -\delta_3, \delta_1)=(0.2, -0.29, 0.09)$ in the triple-$\bm{q}$ SkL phase to satisfy the $\bm{q} \perp \bm{H}$ condition, the Heisenberg exchange energy is only slightly affected. The three $\bm{q}$ vectors in the SkL phase for $\bm{H} \parallel [111]$ simultaneously coincide with the peak in $\chi(\bm{q})$ and drive the SkL formation through the higher-order RKKY interaction~\cite{Hayami21,Hayami17}.

Hereafter, to discuss the helimagnetic structure, we set $\bm{q}=(0.2, 0.3, 0)$ and $S_{\bm{q}}^{(\alpha)}=\{(-3, 2, 0)/\sqrt{13} + (0, 0, i)\}\exp (i \phi_{\alpha})$ as deduced experimentally. We also assume equal magnitudes of moments for Eu-1, 2, 3, and 4. 
With respect to the F-$J_1$ case with $(J_1, J_2)=(1, -0.52)$, the minimum energy ($E=-3.43$) is obtained at approximately $\hat{s}_{\bm{q}}=(1, e^{-i\pi/6}, 1, e^{-i\pi/6})$, where almost the same phase indicates a ferromagnetic local arrangement. 
The real-space helimagnetic structure thus obtained is illustrated in Fig.~\ref{fig:Helical}(a). 
The four spins in a unit cell are oriented in almost the same direction. 
All moments are perpendicular to $\bm{q}$ and rotate counterclockwise when propagating along the $\bm{q}$ vector.

With respect to the AF-$J_1$ case with $(J_1, J_2)=(-1, 0.2)$, the minimum energy ($E=-2.81$) is obtained at approximately $\hat{s}_{\bm{q}}=(1, e^{-i\pi/2}, -1, e^{i\pi/2})$. The sum of the phase factors is zero, indicating an antiferromagnetic local arrangement. 
The real-space helimagnetic structure in this case is illustrated in Fig.~\ref{fig:Helical}(b). 
The four spins in the unit cell are oriented in various directions. 
A serious problem with this structure is that the magnetic structure factor becomes very small due to cancellation among the four Eu sublattices. 
The square of the structure factor, which is proportional to the observed intensity, is two orders of magnitude smaller than that in the F-$J_1$ case in Fig.~\ref{fig:Helical}(a). 
In the SkL structure shown in Fig.~\ref{fig:Fig_SkL}(b), the spins at all four Eu sites are expected to have almost the same phase because they swirl in the same direction. 
Experimentally, the observed peak intensities in the helical phase and SkL phase are almost the same~\cite{Kaneko19,Tabata19}. 
This indicates that the phase relation in the helical phase should be ferromagnetic, that is, the structure shown in Fig.~\ref{fig:Helical}(a) is more likely. 
The observed intensities around the $(\bar{3}, 3, 0)$ fundamental peak also support this conclusion. 
This is also consistent with the predominant ferromagnetic correlation inferred from the positive Weiss temperature of 7.7 K~\cite{Sakakibara19}.

\section{Conclusion}
Using resonant x-ray diffraction with a circularly polarized beam, we studied the magnetic helicity of the triple-$\bm{q}$ magnetic structure of the triangular SkL of EuPtSi for a magnetic field along the [111] axis. 
We demonstrated that all three Fourier components of the triple-$\bm{q}$ structure have the same helicity, which is a direct observation of the SkL in the reciprocal space. By combining the circular polarization dependence and linear polarization analyses, we deduced that the Fourier components are almost circular and perpendicular to the respective $\bm{q}$ vectors. 
Because the helicity of the triple-$\bm{q}$ structure is the same as that of the single-$\bm{q}$ helimagnetic structure at low fields, it is suggested that the antisymmetric exchange interaction inherent in the chiral structure supports the formation of the triangular SkL. We also observed that the helical plane is slightly tilted toward the magnetic field to form a conical structure before the first-order transition to the SkL phase. A possible helimagnetic structure was discussed by considering the nearest and second-nearest neighbor Heisenberg-type exchange interactions. It is suggested that a ferromagnetic nearest-neighbor interaction aligns the four Eu spins in the unit cell in the same direction.

\vspace{5mm}
\acknowledgments
The authors acknowledge valuable comments from T. Nakajima, S. Hayami and Y. Tokunaga. 
This work was supported by JSPS Grant-in-Aid for Scientific Research (B) (No. JP20H01854) and by JSPS Grant-in-Aid for Transformative Research Areas (Asymmetric Quantum Matters, No. JP23H04867). 
The synchrotron experiments were performed under the approval of the Photon Factory Program Advisory Committee (No. 2016G660). 

\appendix
\section{Analysis of $\Delta \theta_{\text{PR}}$ dependence}
We use the scattering-amplitude-operator method to analyze the experimental results of RXD~\cite{Lovesey96}. 
The resonant scattering amplitude can be expressed by a $2\times 2$ matrix $\hat{F}$, consisting of four elements of the scattering amplitude for 
$\sigma$-$\sigma'$, $\pi$-$\sigma'$, $\sigma$-$\pi'$, and $\pi$-$\pi'$:
\begin{equation}
\hat{F} = \begin{pmatrix} F_{\sigma\sigma'} & F_{\pi\sigma'} \\ 
F_{\sigma\pi'} & F_{\pi\pi'} \end{pmatrix}\,.
\label{eq:scampG}
\end{equation}
Using the four elements of (\ref{eq:scampG}), the scattering intensity can be written as 
\begin{align}
I &= 
\frac{1}{2} \bigl(\, |F_{\sigma\sigma'}|^2 + |F_{\sigma\pi'}|^2 + |F_{\pi\sigma'}|^2 + |F_{\pi\pi'}|^2 \,\bigr) \nonumber \\
 &\;\;\;\; + P_1 \text{Re} \bigl\{\, F_{\pi\sigma'}^*F_{\sigma\sigma'} + F_{\pi\pi'}^*F_{\sigma\pi'} \,\bigr\} \nonumber \\
 &\;\;\;\; + P_2 \text{Im} \bigl\{\, F_{\pi\sigma'}^*F_{\sigma\sigma'} + F_{\pi\pi'}^*F_{\sigma\pi'} \,\bigr\} 
 \label{eq:CrossSec1} \\
 &\;\;\;\; +  \frac{1}{2} P_3\bigl(\, |F_{\sigma\sigma'}|^2 + |F_{\sigma\pi'}|^2 - |F_{\pi\sigma'}|^2 - |F_{\pi\pi'}|^2 \,\bigr) 
 \,. \nonumber
\end{align}
Therefore, the intensity for the incident beam described by the Stokes parameters $(P_1, P_2, P_3)$ can generally be written as
\begin{equation}
I = C_0 + C_1 P_1 + C_2 P_2 + C_3 P_3 \,,
\label{eq:CrossSec2}
\end{equation}
which can be used as a fitting function for the $\Delta\theta_{\text{PR}}$ scan with four parameters of $C_n$ ($n=0\sim 3$)~\cite{Matsumura17}.

\begin{figure}[b]
\begin{center}
\includegraphics[width=8cm]{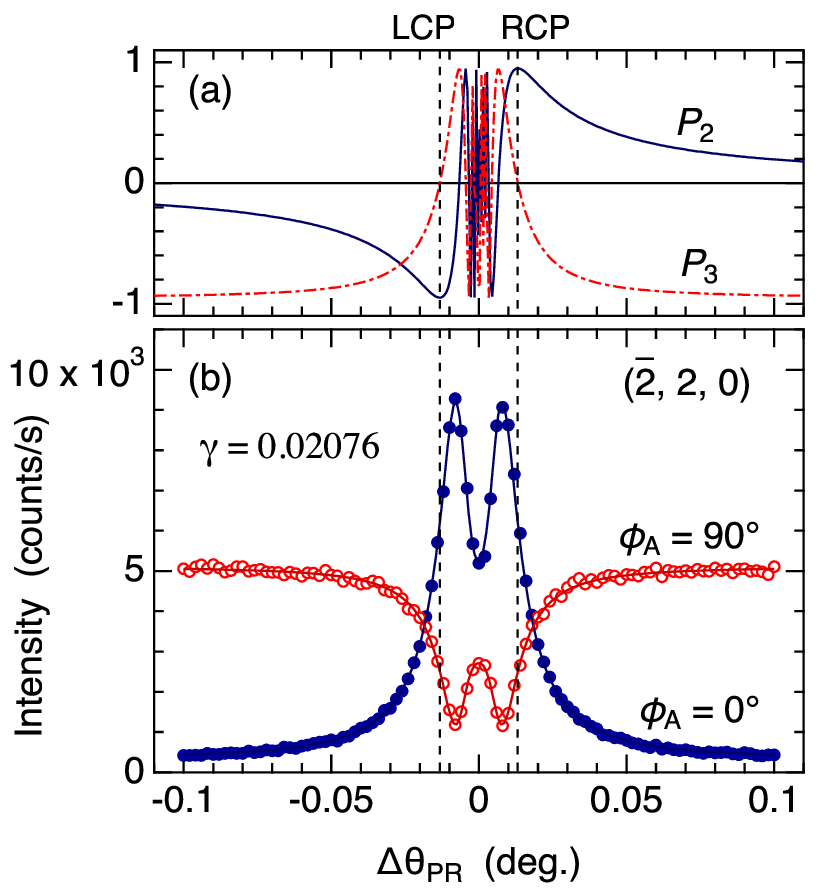} 
\end{center}
\caption{(a) $\Delta \theta_{\text{PR}}$ dependence of the Stokes parameter $P_2$ and $P_3$. 
(b) $\Delta \theta_{\text{PR}}$ dependences of the intensity of the $(\bar{2}, 0, 0)$ fundamental Bragg reflection with linear polarization analysis at $\phi_{\text A}=0^{\circ}$ and $90^{\circ}$. 
Solid lines are the fits, from which $\gamma=0.02076$ was obtained. 
}
\label{fig:FigPR220}
\end{figure}

The $\Delta\theta_{\text{PR}}$ dependence of the Stokes parameter $P_2 = \sin (\gamma/ \Delta \theta_{\text{PR}})$ ($+1$ for RCP and $-1$ for LCP) and $P_3 = -\cos (\gamma/ \Delta \theta_{\text{PR}})$ ($+1$ for $\sigma$ and $-1$ for $\pi$) are shown in Fig.~\ref{fig:FigPR220}(a), 
where $\gamma$ is an experimentally determined parameter of the phase plate obtained by analyzing the $\Delta \theta_{\text{PR}}$ dependence of the intensity of the $(\bar{2}, 2, 0)$ fundamental reflection. 
This is shown in Fig.~\ref{fig:FigPR220}(b). 
The solid lines are the fits with 
\begin{equation}
I = K \Bigl\{ 1 - \frac{(1-P_3)\sin^2 2\theta}{2} \Bigr\} \Bigl\{1 - \frac{(1 - P_{3\text{A}}) \sin^2 2\theta_{\text A}}{2} \Bigr\}
\label{eq:Stokes-30}
\end{equation}
for a nonresonant Thomson scattering, 
where $P_{3\text{A}}$ and $2\theta_{\text{A}}$ are the $P_3$ Stokes parameter and the scattering angle, respectively, at the analyzer crystal. $K$ is a constant scale factor.  
$P_{3\text{A}}$ is expressed as 
\begin{align}
P_{3\text{A}} &= -P_1^{\;\prime} \sin 2\phi_{\text A} + P_3^{\;\prime} \cos 2\phi_{\text A} \;,
\label{eq:Stokes-29}
\end{align}
where $P_1^{\;\prime}$ and $P_3^{\;\prime}$ are the Stokes parameters of the diffracted x ray expressed as 
\begin{align}
P_1^{\;\prime} & = \frac{\displaystyle P_1 \cos 2\theta}{\displaystyle 1 - \frac{1}{2}(1-P_3)\sin^2 2\theta }  \;, \nonumber \\
P_3^{\;\prime} & = \frac{\displaystyle  P_3 + \frac{1}{2} (1-P_3) \sin^2 2\theta }{\displaystyle  1 - \frac{1}{2}(1-P_3)\sin^2 2\theta } \;.
\label{eq:Stokes-29}
\end{align}

\bibliography{EuPtSi-1}

\providecommand{\noopsort}[1]{}\providecommand{\singleletter}[1]{#1}%
\begin{thebibliography}{31}%
\makeatletter
\providecommand \@ifxundefined [1]{%
 \@ifx{#1\undefined}
}%
\providecommand \@ifnum [1]{%
 \ifnum #1\expandafter \@firstoftwo
 \else \expandafter \@secondoftwo
 \fi
}%
\providecommand \@ifx [1]{%
 \ifx #1\expandafter \@firstoftwo
 \else \expandafter \@secondoftwo
 \fi
}%
\providecommand \natexlab [1]{#1}%
\providecommand \enquote  [1]{``#1''}%
\providecommand \bibnamefont  [1]{#1}%
\providecommand \bibfnamefont [1]{#1}%
\providecommand \citenamefont [1]{#1}%
\providecommand \href@noop [0]{\@secondoftwo}%
\providecommand \href [0]{\begingroup \@sanitize@url \@href}%
\providecommand \@href[1]{\@@startlink{#1}\@@href}%
\providecommand \@@href[1]{\endgroup#1\@@endlink}%
\providecommand \@sanitize@url [0]{\catcode `\\12\catcode `\$12\catcode
  `\&12\catcode `\#12\catcode `\^12\catcode `\_12\catcode `\%12\relax}%
\providecommand \@@startlink[1]{}%
\providecommand \@@endlink[0]{}%
\providecommand \url  [0]{\begingroup\@sanitize@url \@url }%
\providecommand \@url [1]{\endgroup\@href {#1}{\urlprefix }}%
\providecommand \urlprefix  [0]{URL }%
\providecommand \Eprint [0]{\href }%
\providecommand \doibase [0]{https://doi.org/}%
\providecommand \selectlanguage [0]{\@gobble}%
\providecommand \bibinfo  [0]{\@secondoftwo}%
\providecommand \bibfield  [0]{\@secondoftwo}%
\providecommand \translation [1]{[#1]}%
\providecommand \BibitemOpen [0]{}%
\providecommand \bibitemStop [0]{}%
\providecommand \bibitemNoStop [0]{.\EOS\space}%
\providecommand \EOS [0]{\spacefactor3000\relax}%
\providecommand \BibitemShut  [1]{\csname bibitem#1\endcsname}%
\let\auto@bib@innerbib\@empty
%</preamble>
\bibitem [{\citenamefont {Kakihana}\ \emph {et~al.}(2018)\citenamefont
  {Kakihana}, \citenamefont {Aoki}, \citenamefont {Nakamura}, \citenamefont
  {Honda}, \citenamefont {Nakashima}, \citenamefont {Amako}, \citenamefont
  {Nakamura}, \citenamefont {Sakakibara}, \citenamefont {Hedo}, \citenamefont
  {Nakama},\ and\ \citenamefont {\={O}nuki}}]{Kakihana18}%
  \BibitemOpen
  \bibfield  {author} {\bibinfo {author} {\bibfnamefont {M.}~\bibnamefont
  {Kakihana}}, \bibinfo {author} {\bibfnamefont {D.}~\bibnamefont {Aoki}},
  \bibinfo {author} {\bibfnamefont {A.}~\bibnamefont {Nakamura}}, \bibinfo
  {author} {\bibfnamefont {F.}~\bibnamefont {Honda}}, \bibinfo {author}
  {\bibfnamefont {M.}~\bibnamefont {Nakashima}}, \bibinfo {author}
  {\bibfnamefont {Y.}~\bibnamefont {Amako}}, \bibinfo {author} {\bibfnamefont
  {S.}~\bibnamefont {Nakamura}}, \bibinfo {author} {\bibfnamefont
  {T.}~\bibnamefont {Sakakibara}}, \bibinfo {author} {\bibfnamefont
  {M.}~\bibnamefont {Hedo}}, \bibinfo {author} {\bibfnamefont {T.}~\bibnamefont
  {Nakama}},\ and\ \bibinfo {author} {\bibfnamefont {Y.}~\bibnamefont
  {\={O}nuki}},\ }\bibfield  {title} {\bibinfo {title} {Giant hall resistivity
  and magnetoresisntance in cubic chiral antiferromagnet {EuPtSi}},\
  }\href@noop {} {\bibfield  {journal} {\bibinfo  {journal} {J. Phys. Soc.
  Jpn.}\ }\textbf {\bibinfo {volume} {87}},\ \bibinfo {pages} {023701}
  (\bibinfo {year} {2018})}\BibitemShut {NoStop}%
\bibitem [{\citenamefont {Kakihana}\ \emph {et~al.}(2019)\citenamefont
  {Kakihana}, \citenamefont {Aoki}, \citenamefont {Nakamura}, \citenamefont
  {Honda}, \citenamefont {Nakashima}, \citenamefont {Amako}, \citenamefont
  {Takeuchi}, \citenamefont {Harima}, \citenamefont {Hedo}, \citenamefont
  {Nakama},\ and\ \citenamefont {\={O}nuki}}]{Kakihana19}%
  \BibitemOpen
  \bibfield  {author} {\bibinfo {author} {\bibfnamefont {M.}~\bibnamefont
  {Kakihana}}, \bibinfo {author} {\bibfnamefont {D.}~\bibnamefont {Aoki}},
  \bibinfo {author} {\bibfnamefont {A.}~\bibnamefont {Nakamura}}, \bibinfo
  {author} {\bibfnamefont {F.}~\bibnamefont {Honda}}, \bibinfo {author}
  {\bibfnamefont {M.}~\bibnamefont {Nakashima}}, \bibinfo {author}
  {\bibfnamefont {Y.}~\bibnamefont {Amako}}, \bibinfo {author} {\bibfnamefont
  {T.}~\bibnamefont {Takeuchi}}, \bibinfo {author} {\bibfnamefont
  {H.}~\bibnamefont {Harima}}, \bibinfo {author} {\bibfnamefont
  {M.}~\bibnamefont {Hedo}}, \bibinfo {author} {\bibfnamefont {T.}~\bibnamefont
  {Nakama}},\ and\ \bibinfo {author} {\bibfnamefont {Y.}~\bibnamefont
  {\={O}nuki}},\ }\bibfield  {title} {\bibinfo {title} {Unique magnetic phases
  in the skyrmion lattice and fermi surface properties in cubic chiral
  antiferromagnet {EuPtSi}},\ }\href@noop {} {\bibfield  {journal} {\bibinfo
  {journal} {J. Phys. Soc. Jpn.}\ }\textbf {\bibinfo {volume} {88}},\ \bibinfo
  {pages} {094705} (\bibinfo {year} {2019})}\BibitemShut {NoStop}%
\bibitem [{\citenamefont {Takeuchi}\ \emph {et~al.}(2019)\citenamefont
  {Takeuchi}, \citenamefont {Kakihana}, \citenamefont {Hedo}, \citenamefont
  {Nakama},\ and\ \citenamefont {\={O}nuki}}]{Takeuchi19}%
  \BibitemOpen
  \bibfield  {author} {\bibinfo {author} {\bibfnamefont {T.}~\bibnamefont
  {Takeuchi}}, \bibinfo {author} {\bibfnamefont {M.}~\bibnamefont {Kakihana}},
  \bibinfo {author} {\bibfnamefont {M.}~\bibnamefont {Hedo}}, \bibinfo {author}
  {\bibfnamefont {T.}~\bibnamefont {Nakama}},\ and\ \bibinfo {author}
  {\bibfnamefont {Y.}~\bibnamefont {\={O}nuki}},\ }\bibfield  {title} {\bibinfo
  {title} {Magnetic field versus temperature phase diagram for {$H \parallel
  [001]$} in the trillium lattice antiferromagnet {EuPtSi}},\ }\href@noop {}
  {\bibfield  {journal} {\bibinfo  {journal} {J. Phys. Soc. Jpn.}\ }\textbf
  {\bibinfo {volume} {88}},\ \bibinfo {pages} {053703} (\bibinfo {year}
  {2019})}\BibitemShut {NoStop}%
\bibitem [{\citenamefont {Sakakibara}\ \emph {et~al.}(2019)\citenamefont
  {Sakakibara}, \citenamefont {Nakamura}, \citenamefont {Kittaka},
  \citenamefont {Kakihana}, \citenamefont {Hedo}, \citenamefont {Nakama},\ and\
  \citenamefont {\={O}nuki}}]{Sakakibara19}%
  \BibitemOpen
  \bibfield  {author} {\bibinfo {author} {\bibfnamefont {T.}~\bibnamefont
  {Sakakibara}}, \bibinfo {author} {\bibfnamefont {S.}~\bibnamefont
  {Nakamura}}, \bibinfo {author} {\bibfnamefont {S.}~\bibnamefont {Kittaka}},
  \bibinfo {author} {\bibfnamefont {M.}~\bibnamefont {Kakihana}}, \bibinfo
  {author} {\bibfnamefont {M.}~\bibnamefont {Hedo}}, \bibinfo {author}
  {\bibfnamefont {T.}~\bibnamefont {Nakama}},\ and\ \bibinfo {author}
  {\bibfnamefont {Y.}~\bibnamefont {\={O}nuki}},\ }\bibfield  {title} {\bibinfo
  {title} {Fluctuation-induced first-order transition and tricritical point in
  {EuPtSi}},\ }\href@noop {} {\bibfield  {journal} {\bibinfo  {journal} {J.
  Phys. Soc. Jpn.}\ }\textbf {\bibinfo {volume} {88}},\ \bibinfo {pages}
  {093701} (\bibinfo {year} {2019})}\BibitemShut {NoStop}%
\bibitem [{\citenamefont {Takeuchi}\ \emph {et~al.}(2020)\citenamefont
  {Takeuchi}, \citenamefont {Kakihana}, \citenamefont {Hedo}, \citenamefont
  {Nakama},\ and\ \citenamefont {\={O}nuki}}]{Takeuchi20}%
  \BibitemOpen
  \bibfield  {author} {\bibinfo {author} {\bibfnamefont {T.}~\bibnamefont
  {Takeuchi}}, \bibinfo {author} {\bibfnamefont {M.}~\bibnamefont {Kakihana}},
  \bibinfo {author} {\bibfnamefont {M.}~\bibnamefont {Hedo}}, \bibinfo {author}
  {\bibfnamefont {T.}~\bibnamefont {Nakama}},\ and\ \bibinfo {author}
  {\bibfnamefont {Y.}~\bibnamefont {\={O}nuki}},\ }\bibfield  {title} {\bibinfo
  {title} {Angle dependence of the magnetic phase diagram in cubic chiral
  antiferromagnet {EuPtSi}},\ }\href@noop {} {\bibfield  {journal} {\bibinfo
  {journal} {J. Phys. Soc. Jpn.}\ }\textbf {\bibinfo {volume} {89}},\ \bibinfo
  {pages} {093703} (\bibinfo {year} {2020})}\BibitemShut {NoStop}%
\bibitem [{\citenamefont {Sakakibara}\ \emph {et~al.}(2021)\citenamefont
  {Sakakibara}, \citenamefont {Nakamura}, \citenamefont {Kittaka},
  \citenamefont {Kakihana}, \citenamefont {Hedo}, \citenamefont {Nakama},\ and\
  \citenamefont {\={O}nuki}}]{Sakakibara21}%
  \BibitemOpen
  \bibfield  {author} {\bibinfo {author} {\bibfnamefont {T.}~\bibnamefont
  {Sakakibara}}, \bibinfo {author} {\bibfnamefont {S.}~\bibnamefont
  {Nakamura}}, \bibinfo {author} {\bibfnamefont {S.}~\bibnamefont {Kittaka}},
  \bibinfo {author} {\bibfnamefont {M.}~\bibnamefont {Kakihana}}, \bibinfo
  {author} {\bibfnamefont {M.}~\bibnamefont {Hedo}}, \bibinfo {author}
  {\bibfnamefont {T.}~\bibnamefont {Nakama}},\ and\ \bibinfo {author}
  {\bibfnamefont {Y.}~\bibnamefont {\={O}nuki}},\ }\bibfield  {title} {\bibinfo
  {title} {Magnetic phase transitions of the {$4f$} skyrmion compound {EuPtSi}
  studied by magnetization measurements},\ }\href@noop {} {\bibfield  {journal}
  {\bibinfo  {journal} {J. Phys. Soc. Jpn.}\ }\textbf {\bibinfo {volume}
  {90}},\ \bibinfo {pages} {064701} (\bibinfo {year} {2021})}\BibitemShut
  {NoStop}%
\bibitem [{\citenamefont {Mishra}\ and\ \citenamefont
  {Ganesan}(2019)}]{Mishra19}%
  \BibitemOpen
  \bibfield  {author} {\bibinfo {author} {\bibfnamefont {A.~K.}\ \bibnamefont
  {Mishra}}\ and\ \bibinfo {author} {\bibfnamefont {V.}~\bibnamefont
  {Ganesan}},\ }\bibfield  {title} {\bibinfo {title} {A-phase, field-induced
  tricritical point, and universal magnetocaloric scaling in {EuPtSi}},\
  }\href@noop {} {\bibfield  {journal} {\bibinfo  {journal} {Phys. Rev. B}\
  }\textbf {\bibinfo {volume} {100}},\ \bibinfo {pages} {125113} (\bibinfo
  {year} {2019})}\BibitemShut {NoStop}%
\bibitem [{\citenamefont {M\"{u}hlbauer}\ \emph {et~al.}(2009)\citenamefont
  {M\"{u}hlbauer}, \citenamefont {Binz}, \citenamefont {Jonietz}, \citenamefont
  {Pfleiderer}, \citenamefont {Rosch}, \citenamefont {Neubauer}, \citenamefont
  {Georgii},\ and\ \citenamefont {B\"{o}ni}}]{Muhlbauer09}%
  \BibitemOpen
  \bibfield  {author} {\bibinfo {author} {\bibfnamefont {S.}~\bibnamefont
  {M\"{u}hlbauer}}, \bibinfo {author} {\bibfnamefont {B.}~\bibnamefont {Binz}},
  \bibinfo {author} {\bibfnamefont {F.}~\bibnamefont {Jonietz}}, \bibinfo
  {author} {\bibfnamefont {C.}~\bibnamefont {Pfleiderer}}, \bibinfo {author}
  {\bibfnamefont {A.}~\bibnamefont {Rosch}}, \bibinfo {author} {\bibfnamefont
  {A.}~\bibnamefont {Neubauer}}, \bibinfo {author} {\bibfnamefont
  {R.}~\bibnamefont {Georgii}},\ and\ \bibinfo {author} {\bibfnamefont
  {P.}~\bibnamefont {B\"{o}ni}},\ }\bibfield  {title} {\bibinfo {title}
  {Skyrmion lattice in a chiral magnet},\ }\href@noop {} {\bibfield  {journal}
  {\bibinfo  {journal} {Science}\ }\textbf {\bibinfo {volume} {323}},\ \bibinfo
  {pages} {915} (\bibinfo {year} {2009})}\BibitemShut {NoStop}%
\bibitem [{\citenamefont {Nagaosa}\ and\ \citenamefont
  {Tokura}(2013)}]{Nagaosa13}%
  \BibitemOpen
  \bibfield  {author} {\bibinfo {author} {\bibfnamefont {N.}~\bibnamefont
  {Nagaosa}}\ and\ \bibinfo {author} {\bibfnamefont {Y.}~\bibnamefont
  {Tokura}},\ }\bibfield  {title} {\bibinfo {title} {Topological properties and
  dynamics of magnetic skyrmions},\ }\href@noop {} {\bibfield  {journal}
  {\bibinfo  {journal} {Nat. Nanotechnol.}\ }\textbf {\bibinfo {volume} {8}},\
  \bibinfo {pages} {899} (\bibinfo {year} {2013})}\BibitemShut {NoStop}%
\bibitem [{\citenamefont {Tokura}\ and\ \citenamefont
  {Kanazawa}(2021)}]{Tokura21}%
  \BibitemOpen
  \bibfield  {author} {\bibinfo {author} {\bibfnamefont {Y.}~\bibnamefont
  {Tokura}}\ and\ \bibinfo {author} {\bibfnamefont {N.}~\bibnamefont
  {Kanazawa}},\ }\bibfield  {title} {\bibinfo {title} {Magnetic skyrmion
  materials},\ }\href@noop {} {\bibfield  {journal} {\bibinfo  {journal} {Chem.
  Rev.}\ }\textbf {\bibinfo {volume} {121}},\ \bibinfo {pages} {2857} (\bibinfo
  {year} {2021})}\BibitemShut {NoStop}%
\bibitem [{\citenamefont {Kaneko}\ \emph {et~al.}(2019)\citenamefont {Kaneko},
  \citenamefont {Frontzek}, \citenamefont {Matsuda}, \citenamefont {Nakao},
  \citenamefont {Munakata}, \citenamefont {Ohhara}, \citenamefont {Kakihana},
  \citenamefont {Haga}, \citenamefont {Hedo}, \citenamefont {Nakama},\ and\
  \citenamefont {\={O}nuki}}]{Kaneko19}%
  \BibitemOpen
  \bibfield  {author} {\bibinfo {author} {\bibfnamefont {K.}~\bibnamefont
  {Kaneko}}, \bibinfo {author} {\bibfnamefont {M.~D.}\ \bibnamefont
  {Frontzek}}, \bibinfo {author} {\bibfnamefont {M.}~\bibnamefont {Matsuda}},
  \bibinfo {author} {\bibfnamefont {A.}~\bibnamefont {Nakao}}, \bibinfo
  {author} {\bibfnamefont {K.}~\bibnamefont {Munakata}}, \bibinfo {author}
  {\bibfnamefont {T.}~\bibnamefont {Ohhara}}, \bibinfo {author} {\bibfnamefont
  {M.}~\bibnamefont {Kakihana}}, \bibinfo {author} {\bibfnamefont
  {Y.}~\bibnamefont {Haga}}, \bibinfo {author} {\bibfnamefont {M.}~\bibnamefont
  {Hedo}}, \bibinfo {author} {\bibfnamefont {T.}~\bibnamefont {Nakama}},\ and\
  \bibinfo {author} {\bibfnamefont {Y.}~\bibnamefont {\={O}nuki}},\ }\bibfield
  {title} {\bibinfo {title} {Unique helical magnetic order and field-induced
  phase in trillium lattice antiferromagnet {EuPtSi}},\ }\href@noop {}
  {\bibfield  {journal} {\bibinfo  {journal} {J. Phys. Soc. Jpn.}\ }\textbf
  {\bibinfo {volume} {88}},\ \bibinfo {pages} {013702} (\bibinfo {year}
  {2019})}\BibitemShut {NoStop}%
\bibitem [{\citenamefont {Tabata}\ \emph {et~al.}(2019)\citenamefont {Tabata},
  \citenamefont {Matsumura}, \citenamefont {Nakao}, \citenamefont {Michimura},
  \citenamefont {Kakihana}, \citenamefont {Inami}, \citenamefont {Kaneko},
  \citenamefont {Hedo}, \citenamefont {Nakama},\ and\ \citenamefont
  {\={O}nuki}}]{Tabata19}%
  \BibitemOpen
  \bibfield  {author} {\bibinfo {author} {\bibfnamefont {C.}~\bibnamefont
  {Tabata}}, \bibinfo {author} {\bibfnamefont {T.}~\bibnamefont {Matsumura}},
  \bibinfo {author} {\bibfnamefont {H.}~\bibnamefont {Nakao}}, \bibinfo
  {author} {\bibfnamefont {S.}~\bibnamefont {Michimura}}, \bibinfo {author}
  {\bibfnamefont {M.}~\bibnamefont {Kakihana}}, \bibinfo {author}
  {\bibfnamefont {T.}~\bibnamefont {Inami}}, \bibinfo {author} {\bibfnamefont
  {K.}~\bibnamefont {Kaneko}}, \bibinfo {author} {\bibfnamefont
  {M.}~\bibnamefont {Hedo}}, \bibinfo {author} {\bibfnamefont {T.}~\bibnamefont
  {Nakama}},\ and\ \bibinfo {author} {\bibfnamefont {Y.}~\bibnamefont
  {\={O}nuki}},\ }\bibfield  {title} {\bibinfo {title} {Magnetic field induced
  triple-q magnetic order in trillium lattice antiferromagnet {EuPtSi} studied
  by resonant x-ray scattering},\ }\href@noop {} {\bibfield  {journal}
  {\bibinfo  {journal} {J. Phys. Soc. Jpn.}\ }\textbf {\bibinfo {volume}
  {88}},\ \bibinfo {pages} {093704} (\bibinfo {year} {2019})}\BibitemShut
  {NoStop}%
\bibitem [{\citenamefont {Yu}\ \emph {et~al.}(2010)\citenamefont {Yu},
  \citenamefont {Onose}, \citenamefont {Kanazawa}, \citenamefont {Park},
  \citenamefont {Han}, \citenamefont {Matsui}, \citenamefont {Nagaosa},\ and\
  \citenamefont {Tokura}}]{Yu10}%
  \BibitemOpen
  \bibfield  {author} {\bibinfo {author} {\bibfnamefont {X.~Z.}\ \bibnamefont
  {Yu}}, \bibinfo {author} {\bibfnamefont {Y.}~\bibnamefont {Onose}}, \bibinfo
  {author} {\bibfnamefont {N.}~\bibnamefont {Kanazawa}}, \bibinfo {author}
  {\bibfnamefont {J.~H.}\ \bibnamefont {Park}}, \bibinfo {author}
  {\bibfnamefont {J.~H.}\ \bibnamefont {Han}}, \bibinfo {author} {\bibfnamefont
  {Y.}~\bibnamefont {Matsui}}, \bibinfo {author} {\bibfnamefont
  {N.}~\bibnamefont {Nagaosa}},\ and\ \bibinfo {author} {\bibfnamefont
  {Y.}~\bibnamefont {Tokura}},\ }\bibfield  {title} {\bibinfo {title}
  {Real-space observation of a two-dimensional skyrmion crystal},\ }\href@noop
  {} {\bibfield  {journal} {\bibinfo  {journal} {Nature}\ }\textbf {\bibinfo
  {volume} {465}},\ \bibinfo {pages} {901} (\bibinfo {year}
  {2010})}\BibitemShut {NoStop}%
\bibitem [{\citenamefont {Khanh}\ \emph {et~al.}(2020)\citenamefont {Khanh},
  \citenamefont {Nakajima}, \citenamefont {Yu}, \citenamefont {Gao},
  \citenamefont {Shibata}, \citenamefont {Hirschberger}, \citenamefont
  {Yamasaki}, \citenamefont {Sagayama}, \citenamefont {Nakao}, \citenamefont
  {Peng}, \citenamefont {Nakajima}, \citenamefont {Takagi}, \citenamefont
  {Arima}, \citenamefont {Tokura},\ and\ \citenamefont {Seki}}]{Khanh20}%
  \BibitemOpen
  \bibfield  {author} {\bibinfo {author} {\bibfnamefont {N.~D.}\ \bibnamefont
  {Khanh}}, \bibinfo {author} {\bibfnamefont {T.}~\bibnamefont {Nakajima}},
  \bibinfo {author} {\bibfnamefont {X.}~\bibnamefont {Yu}}, \bibinfo {author}
  {\bibfnamefont {S.}~\bibnamefont {Gao}}, \bibinfo {author} {\bibfnamefont
  {K.}~\bibnamefont {Shibata}}, \bibinfo {author} {\bibfnamefont
  {M.}~\bibnamefont {Hirschberger}}, \bibinfo {author} {\bibfnamefont
  {Y.}~\bibnamefont {Yamasaki}}, \bibinfo {author} {\bibfnamefont
  {H.}~\bibnamefont {Sagayama}}, \bibinfo {author} {\bibfnamefont
  {H.}~\bibnamefont {Nakao}}, \bibinfo {author} {\bibfnamefont
  {L.}~\bibnamefont {Peng}}, \bibinfo {author} {\bibfnamefont {K.}~\bibnamefont
  {Nakajima}}, \bibinfo {author} {\bibfnamefont {R.}~\bibnamefont {Takagi}},
  \bibinfo {author} {\bibfnamefont {T.}~\bibnamefont {Arima}}, \bibinfo
  {author} {\bibfnamefont {Y.}~\bibnamefont {Tokura}},\ and\ \bibinfo {author}
  {\bibfnamefont {S.}~\bibnamefont {Seki}},\ }\bibfield  {title} {\bibinfo
  {title} {Nanometric square skyrmion lattice in a centrosymmetric tetragonal
  magnet},\ }\href@noop {} {\bibfield  {journal} {\bibinfo  {journal} {Nat.
  Nanotechnol.}\ }\textbf {\bibinfo {volume} {15}},\ \bibinfo {pages} {444}
  (\bibinfo {year} {2020})}\BibitemShut {NoStop}%
\bibitem [{\citenamefont {Hopkinson}\ and\ \citenamefont
  {Kee}(2006)}]{Hopkinson06}%
  \BibitemOpen
  \bibfield  {author} {\bibinfo {author} {\bibfnamefont {J.~M.}\ \bibnamefont
  {Hopkinson}}\ and\ \bibinfo {author} {\bibfnamefont {H.-Y.}\ \bibnamefont
  {Kee}},\ }\bibfield  {title} {\bibinfo {title} {Geometric frustration
  inherent to the trillium lattice, a sublattice of the {B20} structure},\
  }\href@noop {} {\bibfield  {journal} {\bibinfo  {journal} {Phys. Rev. B}\
  }\textbf {\bibinfo {volume} {74}},\ \bibinfo {pages} {224441} (\bibinfo
  {year} {2006})}\BibitemShut {NoStop}%
\bibitem [{\citenamefont {Isakov}\ \emph {et~al.}(2008)\citenamefont {Isakov},
  \citenamefont {Hopkinson},\ and\ \citenamefont {Kee}}]{Isakov08}%
  \BibitemOpen
  \bibfield  {author} {\bibinfo {author} {\bibfnamefont {S.~V.}\ \bibnamefont
  {Isakov}}, \bibinfo {author} {\bibfnamefont {J.~M.}\ \bibnamefont
  {Hopkinson}},\ and\ \bibinfo {author} {\bibfnamefont {H.-Y.}\ \bibnamefont
  {Kee}},\ }\bibfield  {title} {\bibinfo {title} {Fate of partial order on
  trillium and distorted widnmill lattices},\ }\href@noop {} {\bibfield
  {journal} {\bibinfo  {journal} {Phys. Rev. B}\ }\textbf {\bibinfo {volume}
  {78}},\ \bibinfo {pages} {014404} (\bibinfo {year} {2008})}\BibitemShut
  {NoStop}%
\bibitem [{\citenamefont {Redpath}\ and\ \citenamefont
  {Hopkinson}(2010)}]{Redpath10}%
  \BibitemOpen
  \bibfield  {author} {\bibinfo {author} {\bibfnamefont {T.~E.}\ \bibnamefont
  {Redpath}}\ and\ \bibinfo {author} {\bibfnamefont {J.~M.}\ \bibnamefont
  {Hopkinson}},\ }\bibfield  {title} {\bibinfo {title} {Spin ice on the
  trillium lattice studied by {Monte Carlo} calculations},\ }\href@noop {}
  {\bibfield  {journal} {\bibinfo  {journal} {Phys. Rev. B}\ }\textbf {\bibinfo
  {volume} {82}},\ \bibinfo {pages} {014410} (\bibinfo {year}
  {2010})}\BibitemShut {NoStop}%
\bibitem [{\citenamefont {Franco}\ \emph {et~al.}(2017)\citenamefont {Franco},
  \citenamefont {Prots}, \citenamefont {Geibel},\ and\ \citenamefont
  {Seiro}}]{Franco17}%
  \BibitemOpen
  \bibfield  {author} {\bibinfo {author} {\bibfnamefont {D.~G.}\ \bibnamefont
  {Franco}}, \bibinfo {author} {\bibfnamefont {Y.}~\bibnamefont {Prots}},
  \bibinfo {author} {\bibfnamefont {C.}~\bibnamefont {Geibel}},\ and\ \bibinfo
  {author} {\bibfnamefont {S.}~\bibnamefont {Seiro}},\ }\bibfield  {title}
  {\bibinfo {title} {Fluctuation-induced first-order transition in {Eu}-based
  trillium lattices},\ }\href@noop {} {\bibfield  {journal} {\bibinfo
  {journal} {Phys. Rev. B}\ }\textbf {\bibinfo {volume} {96}},\ \bibinfo
  {pages} {014401} (\bibinfo {year} {2017})}\BibitemShut {NoStop}%
\bibitem [{\citenamefont {Homma}\ \emph {et~al.}(2019)\citenamefont {Homma},
  \citenamefont {Kakihana}, \citenamefont {Tokunaga}, \citenamefont {Yogi},
  \citenamefont {Nakashima}, \citenamefont {Nakamura}, \citenamefont {Shimizu},
  \citenamefont {Li}, \citenamefont {Maurya}, \citenamefont {Sato},
  \citenamefont {Honda}, \citenamefont {Aoki}, \citenamefont {Amako},
  \citenamefont {Hedo}, \citenamefont {Nakama},\ and\ \citenamefont
  {\={O}nuki}}]{Homma19}%
  \BibitemOpen
  \bibfield  {author} {\bibinfo {author} {\bibfnamefont {Y.}~\bibnamefont
  {Homma}}, \bibinfo {author} {\bibfnamefont {M.}~\bibnamefont {Kakihana}},
  \bibinfo {author} {\bibfnamefont {Y.}~\bibnamefont {Tokunaga}}, \bibinfo
  {author} {\bibfnamefont {M.}~\bibnamefont {Yogi}}, \bibinfo {author}
  {\bibfnamefont {M.}~\bibnamefont {Nakashima}}, \bibinfo {author}
  {\bibfnamefont {A.}~\bibnamefont {Nakamura}}, \bibinfo {author}
  {\bibfnamefont {Y.}~\bibnamefont {Shimizu}}, \bibinfo {author} {\bibfnamefont
  {D.~X.}\ \bibnamefont {Li}}, \bibinfo {author} {\bibfnamefont
  {A.}~\bibnamefont {Maurya}}, \bibinfo {author} {\bibfnamefont {Y.~J.}\
  \bibnamefont {Sato}}, \bibinfo {author} {\bibfnamefont {F.}~\bibnamefont
  {Honda}}, \bibinfo {author} {\bibfnamefont {D.}~\bibnamefont {Aoki}},
  \bibinfo {author} {\bibfnamefont {Y.}~\bibnamefont {Amako}}, \bibinfo
  {author} {\bibfnamefont {M.}~\bibnamefont {Hedo}}, \bibinfo {author}
  {\bibfnamefont {T.}~\bibnamefont {Nakama}},\ and\ \bibinfo {author}
  {\bibfnamefont {Y.}~\bibnamefont {\={O}nuki}},\ }\bibfield  {title} {\bibinfo
  {title} {Magnetic fluctuation and first-order transition in trillium lattice
  of {EuPtSi} observed by {$^{151}$Eu} {M\"ossbauer} spectroscopy},\
  }\href@noop {} {\bibfield  {journal} {\bibinfo  {journal} {J. Phys. Soc.
  Jpn.}\ }\textbf {\bibinfo {volume} {88}},\ \bibinfo {pages} {094702}
  (\bibinfo {year} {2019})}\BibitemShut {NoStop}%
\bibitem [{\citenamefont {Higa}\ \emph {et~al.}(2021)\citenamefont {Higa},
  \citenamefont {Ito}, \citenamefont {Yogi}, \citenamefont {Hattori},
  \citenamefont {Sakai}, \citenamefont {Kambe}, \citenamefont {Guguchia},
  \citenamefont {Higemoto}, \citenamefont {Nakashima}, \citenamefont {Homma},
  \citenamefont {Nakamura}, \citenamefont {Honda}, \citenamefont {Shimizu},
  \citenamefont {Aoki}, \citenamefont {Kakihana}, \citenamefont {Hedo},
  \citenamefont {Nakama}, \citenamefont {\={O}nuki},\ and\ \citenamefont
  {Tokunaga}}]{Higa21}%
  \BibitemOpen
  \bibfield  {author} {\bibinfo {author} {\bibfnamefont {N.}~\bibnamefont
  {Higa}}, \bibinfo {author} {\bibfnamefont {T.~U.}\ \bibnamefont {Ito}},
  \bibinfo {author} {\bibfnamefont {M.}~\bibnamefont {Yogi}}, \bibinfo {author}
  {\bibfnamefont {T.}~\bibnamefont {Hattori}}, \bibinfo {author} {\bibfnamefont
  {H.}~\bibnamefont {Sakai}}, \bibinfo {author} {\bibfnamefont
  {S.}~\bibnamefont {Kambe}}, \bibinfo {author} {\bibfnamefont
  {Z.}~\bibnamefont {Guguchia}}, \bibinfo {author} {\bibfnamefont
  {W.}~\bibnamefont {Higemoto}}, \bibinfo {author} {\bibfnamefont
  {M.}~\bibnamefont {Nakashima}}, \bibinfo {author} {\bibfnamefont
  {Y.}~\bibnamefont {Homma}}, \bibinfo {author} {\bibfnamefont
  {A.}~\bibnamefont {Nakamura}}, \bibinfo {author} {\bibfnamefont
  {F.}~\bibnamefont {Honda}}, \bibinfo {author} {\bibfnamefont
  {Y.}~\bibnamefont {Shimizu}}, \bibinfo {author} {\bibfnamefont
  {D.}~\bibnamefont {Aoki}}, \bibinfo {author} {\bibfnamefont {M.}~\bibnamefont
  {Kakihana}}, \bibinfo {author} {\bibfnamefont {M.}~\bibnamefont {Hedo}},
  \bibinfo {author} {\bibfnamefont {T.}~\bibnamefont {Nakama}}, \bibinfo
  {author} {\bibfnamefont {Y.}~\bibnamefont {\={O}nuki}},\ and\ \bibinfo
  {author} {\bibfnamefont {Y.}~\bibnamefont {Tokunaga}},\ }\bibfield  {title}
  {\bibinfo {title} {Critical slowing-down and field-dependent paramagnetic
  fluctuations in the skyrmion host {EuPtSi}: {$\mu$SR} and {NMR} studies},\
  }\href@noop {} {\bibfield  {journal} {\bibinfo  {journal} {Phys. Rev. B}\
  }\textbf {\bibinfo {volume} {104}},\ \bibinfo {pages} {045145} (\bibinfo
  {year} {2021})}\BibitemShut {NoStop}%
\bibitem [{\citenamefont {Yambe}\ and\ \citenamefont {Hayami}(2022)}]{Yambe22}%
  \BibitemOpen
  \bibfield  {author} {\bibinfo {author} {\bibfnamefont {R.}~\bibnamefont
  {Yambe}}\ and\ \bibinfo {author} {\bibfnamefont {S.}~\bibnamefont {Hayami}},\
  }\bibfield  {title} {\bibinfo {title} {Effective spin model in momentum
  space: Toward a systematic understanding of multiple-{$Q$} instability by
  momentum-resolved anisotropic exchange interactions},\ }\href@noop {}
  {\bibfield  {journal} {\bibinfo  {journal} {Phys. Rev. B}\ }\textbf {\bibinfo
  {volume} {106}},\ \bibinfo {pages} {174437} (\bibinfo {year}
  {2022})}\BibitemShut {NoStop}%
\bibitem [{\citenamefont {Hayami}\ and\ \citenamefont
  {Motome}(2018)}]{Hayami18}%
  \BibitemOpen
  \bibfield  {author} {\bibinfo {author} {\bibfnamefont {S.}~\bibnamefont
  {Hayami}}\ and\ \bibinfo {author} {\bibfnamefont {Y.}~\bibnamefont
  {Motome}},\ }\bibfield  {title} {\bibinfo {title} {{N\'{e}el}- and
  {Bloch}-type magnetic vortices in {Rashba} metals},\ }\href@noop {}
  {\bibfield  {journal} {\bibinfo  {journal} {Phys. Rev. Lett.}\ }\textbf
  {\bibinfo {volume} {121}},\ \bibinfo {pages} {137202} (\bibinfo {year}
  {2018})}\BibitemShut {NoStop}%
\bibitem [{\citenamefont {Hayami}\ and\ \citenamefont
  {Motome}(2021)}]{Hayami21b}%
  \BibitemOpen
  \bibfield  {author} {\bibinfo {author} {\bibfnamefont {S.}~\bibnamefont
  {Hayami}}\ and\ \bibinfo {author} {\bibfnamefont {Y.}~\bibnamefont
  {Motome}},\ }\bibfield  {title} {\bibinfo {title} {Noncoplanar multiple-{$Q$}
  spin textures by itinerant frustration: Effects of single-ion anisotropy and
  bond-dependent anisotropy},\ }\href@noop {} {\bibfield  {journal} {\bibinfo
  {journal} {Phys. Rev. B}\ }\textbf {\bibinfo {volume} {103}},\ \bibinfo
  {pages} {054422} (\bibinfo {year} {2021})}\BibitemShut {NoStop}%
\bibitem [{\citenamefont {Hayami}\ and\ \citenamefont
  {Yambe}(2021)}]{Hayami21}%
  \BibitemOpen
  \bibfield  {author} {\bibinfo {author} {\bibfnamefont {S.}~\bibnamefont
  {Hayami}}\ and\ \bibinfo {author} {\bibfnamefont {R.}~\bibnamefont {Yambe}},\
  }\bibfield  {title} {\bibinfo {title} {Field-direction sensitive skyrmion
  crystals in cubic chiral systems: Implication to 4$f$-electron compound
  {EuPtSi}},\ }\href@noop {} {\bibfield  {journal} {\bibinfo  {journal} {J.
  Phys. Soc. Jpn.}\ }\textbf {\bibinfo {volume} {90}},\ \bibinfo {pages}
  {073705} (\bibinfo {year} {2021})}\BibitemShut {NoStop}%
\bibitem [{\citenamefont {Hayami}\ \emph {et~al.}(2017)\citenamefont {Hayami},
  \citenamefont {Ozawa},\ and\ \citenamefont {Motome}}]{Hayami17}%
  \BibitemOpen
  \bibfield  {author} {\bibinfo {author} {\bibfnamefont {S.}~\bibnamefont
  {Hayami}}, \bibinfo {author} {\bibfnamefont {R.}~\bibnamefont {Ozawa}},\ and\
  \bibinfo {author} {\bibfnamefont {Y.}~\bibnamefont {Motome}},\ }\bibfield
  {title} {\bibinfo {title} {Effective bilinear-biquadratic model for
  noncoplanar ordering in itinerant magnets},\ }\href@noop {} {\bibfield
  {journal} {\bibinfo  {journal} {Phys. Rev. B}\ }\textbf {\bibinfo {volume}
  {95}},\ \bibinfo {pages} {224424} (\bibinfo {year} {2017})}\BibitemShut
  {NoStop}%
\bibitem [{\citenamefont {Matsumura}\ \emph {et~al.}(2023)\citenamefont
  {Matsumura}, \citenamefont {Kurauchi}, \citenamefont {Tsukagoshi},
  \citenamefont {Higa}, \citenamefont {Nakao}, \citenamefont {Kakihana},
  \citenamefont {Hedo}, \citenamefont {Nakama},\ and\ \citenamefont
  {\={O}nuki}}]{Matsumura23}%
  \BibitemOpen
  \bibfield  {author} {\bibinfo {author} {\bibfnamefont {T.}~\bibnamefont
  {Matsumura}}, \bibinfo {author} {\bibfnamefont {K.}~\bibnamefont {Kurauchi}},
  \bibinfo {author} {\bibfnamefont {M.}~\bibnamefont {Tsukagoshi}}, \bibinfo
  {author} {\bibfnamefont {N.}~\bibnamefont {Higa}}, \bibinfo {author}
  {\bibfnamefont {H.}~\bibnamefont {Nakao}}, \bibinfo {author} {\bibfnamefont
  {M.}~\bibnamefont {Kakihana}}, \bibinfo {author} {\bibfnamefont
  {M.}~\bibnamefont {Hedo}}, \bibinfo {author} {\bibfnamefont {T.}~\bibnamefont
  {Nakama}},\ and\ \bibinfo {author} {\bibfnamefont {Y.}~\bibnamefont
  {\={O}nuki}},\ }\href@noop {} {\bibfield  {journal} {\bibinfo  {journal}
  {arXiv.2306.14767}\ } (\bibinfo {year} {2023})}\BibitemShut {NoStop}%
\bibitem [{\citenamefont {Shimizu}\ \emph {et~al.}(2022)\citenamefont
  {Shimizu}, \citenamefont {Okumura}, \citenamefont {Kato},\ and\ \citenamefont
  {Motome}}]{Shimizu22}%
  \BibitemOpen
  \bibfield  {author} {\bibinfo {author} {\bibfnamefont {K.}~\bibnamefont
  {Shimizu}}, \bibinfo {author} {\bibfnamefont {S.}~\bibnamefont {Okumura}},
  \bibinfo {author} {\bibfnamefont {Y.}~\bibnamefont {Kato}},\ and\ \bibinfo
  {author} {\bibfnamefont {Y.}~\bibnamefont {Motome}},\ }\bibfield  {title}
  {\bibinfo {title} {Phase degree of freedom and topology in multiple-{$Q$}
  spin textures},\ }\href@noop {} {\bibfield  {journal} {\bibinfo  {journal}
  {Phys. Rev. B}\ }\textbf {\bibinfo {volume} {105}},\ \bibinfo {pages}
  {224405} (\bibinfo {year} {2022})}\BibitemShut {NoStop}%
\bibitem [{\citenamefont {Janoschek}\ \emph {et~al.}(2013)\citenamefont
  {Janoschek}, \citenamefont {Garst}, \citenamefont {Bauer}, \citenamefont
  {Krautscheid}, \citenamefont {Georgii}, \citenamefont {B\"{o}ni},\ and\
  \citenamefont {Pfleiderer}}]{Janoschek13}%
  \BibitemOpen
  \bibfield  {author} {\bibinfo {author} {\bibfnamefont {M.}~\bibnamefont
  {Janoschek}}, \bibinfo {author} {\bibfnamefont {M.}~\bibnamefont {Garst}},
  \bibinfo {author} {\bibfnamefont {A.}~\bibnamefont {Bauer}}, \bibinfo
  {author} {\bibfnamefont {P.}~\bibnamefont {Krautscheid}}, \bibinfo {author}
  {\bibfnamefont {R.}~\bibnamefont {Georgii}}, \bibinfo {author} {\bibfnamefont
  {P.}~\bibnamefont {B\"{o}ni}},\ and\ \bibinfo {author} {\bibfnamefont
  {C.}~\bibnamefont {Pfleiderer}},\ }\bibfield  {title} {\bibinfo {title}
  {Fluctuation-induced first-order phase transition in {Dzyaloshinskii-Moriya}
  helimagnets},\ }\href@noop {} {\bibfield  {journal} {\bibinfo  {journal}
  {Phys. Rev. B}\ }\textbf {\bibinfo {volume} {87}},\ \bibinfo {pages} {134407}
  (\bibinfo {year} {2013})}\BibitemShut {NoStop}%
\bibitem [{\citenamefont {Pappas}\ \emph {et~al.}(2017)\citenamefont {Pappas},
  \citenamefont {Bannenberg}, \citenamefont {Leli\`{e}vre-Berna}, \citenamefont
  {Qian}, \citenamefont {Dewhurst}, \citenamefont {Dalgliesh}, \citenamefont
  {Schlagel}, \citenamefont {Lograsso},\ and\ \citenamefont
  {Falus}}]{Pappas17}%
  \BibitemOpen
  \bibfield  {author} {\bibinfo {author} {\bibfnamefont {C.}~\bibnamefont
  {Pappas}}, \bibinfo {author} {\bibfnamefont {L.~J.}\ \bibnamefont
  {Bannenberg}}, \bibinfo {author} {\bibfnamefont {E.}~\bibnamefont
  {Leli\`{e}vre-Berna}}, \bibinfo {author} {\bibfnamefont {F.}~\bibnamefont
  {Qian}}, \bibinfo {author} {\bibfnamefont {C.~D.}\ \bibnamefont {Dewhurst}},
  \bibinfo {author} {\bibfnamefont {R.~M.}\ \bibnamefont {Dalgliesh}}, \bibinfo
  {author} {\bibfnamefont {D.~L.}\ \bibnamefont {Schlagel}}, \bibinfo {author}
  {\bibfnamefont {T.~A.}\ \bibnamefont {Lograsso}},\ and\ \bibinfo {author}
  {\bibfnamefont {P.}~\bibnamefont {Falus}},\ }\bibfield  {title} {\bibinfo
  {title} {Magnetic fluctuations, precursor phenomena, and phase transition in
  {MnSi} under a magnetic field},\ }\href@noop {} {\bibfield  {journal}
  {\bibinfo  {journal} {Phys. Rev. Lett.}\ }\textbf {\bibinfo {volume} {119}},\
  \bibinfo {pages} {047203} (\bibinfo {year} {2017})}\BibitemShut {NoStop}%
\bibitem [{\citenamefont {Lovesey}\ and\ \citenamefont
  {Collins}(1996)}]{Lovesey96}%
  \BibitemOpen
  \bibfield  {author} {\bibinfo {author} {\bibfnamefont {S.~W.}\ \bibnamefont
  {Lovesey}}\ and\ \bibinfo {author} {\bibfnamefont {S.~P.}\ \bibnamefont
  {Collins}},\ }\href@noop {} {\emph {\bibinfo {title} {X-ray Scattering and
  Absorption by Magnetic Materials}}}\ (\bibinfo  {publisher} {Oxford},\
  \bibinfo {address} {New York},\ \bibinfo {year} {1996})\BibitemShut {NoStop}%
\bibitem [{\citenamefont {Matsumura}\ \emph {et~al.}(2017)\citenamefont
  {Matsumura}, \citenamefont {Kita}, \citenamefont {Kubo}, \citenamefont
  {Yoshikawa}, \citenamefont {Michimura}, \citenamefont {Inami}, \citenamefont
  {Kousaka}, \citenamefont {Inoue},\ and\ \citenamefont {Ohara}}]{Matsumura17}%
  \BibitemOpen
  \bibfield  {author} {\bibinfo {author} {\bibfnamefont {T.}~\bibnamefont
  {Matsumura}}, \bibinfo {author} {\bibfnamefont {Y.}~\bibnamefont {Kita}},
  \bibinfo {author} {\bibfnamefont {K.}~\bibnamefont {Kubo}}, \bibinfo {author}
  {\bibfnamefont {Y.}~\bibnamefont {Yoshikawa}}, \bibinfo {author}
  {\bibfnamefont {S.}~\bibnamefont {Michimura}}, \bibinfo {author}
  {\bibfnamefont {T.}~\bibnamefont {Inami}}, \bibinfo {author} {\bibfnamefont
  {Y.}~\bibnamefont {Kousaka}}, \bibinfo {author} {\bibfnamefont
  {K.}~\bibnamefont {Inoue}},\ and\ \bibinfo {author} {\bibfnamefont
  {S.}~\bibnamefont {Ohara}},\ }\bibfield  {title} {\bibinfo {title} {Chiral
  soliton lattice formation in monoaxial helimagnet
  {Yb(Ni$_{1-x}$Cu$_x$)$_3$Al$_9$}},\ }\href@noop {} {\bibfield  {journal}
  {\bibinfo  {journal} {J. Phys. Soc. Jpn.}\ }\textbf {\bibinfo {volume}
  {86}},\ \bibinfo {pages} {124702} (\bibinfo {year} {2017})}\BibitemShut
  {NoStop}%
\end{thebibliography}%

\end{document}